\documentclass[reprint,eqsecnum,floats,aps,amsmath,amssymb,nofootinbib,prd,onecolumn, showpacs]{revtex4-1}

\usepackage{graphicx}
\usepackage{amsmath,amssymb}
\usepackage{hyperref}
\usepackage{graphicx}
\usepackage{subfigure}
\usepackage{arydshln}

\begin{document}
	
	\title{Backreaction of fermionic perturbations in the Hamiltonian of hybrid loop quantum cosmology}
	
	\author{Beatriz Elizaga Navascu\'es}
	\email{beatriz.b.elizaga@gravity.fau.de}
	\affiliation{Institute for Quantum Gravity, Friedrich-Alexander University Erlangen-N{\"u}rnberg, Staudstra{\ss}e 7, 91058 Erlangen, Germany}
	\author{Guillermo  A. Mena Marug\'an}
	\email{mena@iem.cfmac.csic.es}
	\affiliation{Instituto de Estructura de la Materia, IEM-CSIC, Serrano 121, 28006 Madrid, Spain}
	\author{Santiago Prado Loy}
	\email{santiago.prado@iem.cfmac.csic.es}
	\affiliation{Instituto de Estructura de la Materia, IEM-CSIC, Serrano 121, 28006 Madrid, Spain}

\begin{abstract}
		
We discuss the freedom available in hybrid loop quantum cosmology to define canonical variables for the matter content and investigate whether this can be used to derive a quantum field theory with good properties for the matter sector. We study a primordial, inflationary, cosmological spacetime with inhomogeneous perturbations at lowest nontrivial order, and focus our attention on the contribution of minimally coupled fermionic perturbations of Dirac type. Within the framework of the hybrid quantization, we analyze the different possible separations of the homogeneous background and the inhomogeneous perturbations, by means of canonical transformations that mix the two separated sectors. These possibilities provide a family of sets of annihilation and creationlike fermionic variables, each of them with a different associated contribution to the total Hamiltonian. In all cases, imposing the quantum constraints and introducing a Born-Oppenheimer approximation, one can derive a Schr\"odinger equation for the fermionic part of the wave functions. The resulting evolution turns out to be generated, for each of the allowed choices of variables, by a version of the fermionic contribution to the Hamiltonian which is obtained by evaluating all the dependence on the homogeneous geometry at quantum expectation values. This equation contains a term that encodes the backreaction of the fermionic perturbations on the quantum dynamics of the homogeneous sector. We analyze this backreaction by solving the associated Heisenberg evolution of the fermionic annihilation and creation operators. Then, we identify the conditions that the choice of those operators must satisfy in order to lead to a finite backreaction. Finally, we discuss further restrictions on this choice so that the fermionic Hamiltonian that dictates the Schr\"odinger dynamics is densely defined in Fock space.
\end{abstract}


\maketitle

\section{Introduction}

In conventional quantum theories of matter fields, one employs, in one way or another, some type of renormalization or regularization procedure to obtain physically acceptable results. Such techniques are especially well understood when it comes to (perturbatively) describing the nongravitational interactions contained in the standard model of particle physics. Nonetheless, the issue exceeds this traditional framework in high energy physics. Actually, divergences become even more severe when one considers matter fields propagating in generally curved spacetimes, as it is allowed by Einstein's theory. In those cases, one usually considers that the matter fields are coupled gravitationally to the spacetime, which is viewed as a classical entity. Besides, one frequently neglects the contribution of the fields to the dynamics of the spacetime geometry itself. In such scenarios, infinities generically arise in the quantum theories that describe the matter fields. This problem has been studied in depth over the last decades (see, e.g., Refs. \cite{christ1,christ2,alder,wald1,wald2,wald3,wald4,wald5,wald,bidav,wfs}), and it is commonly believed that the reasons behind it can be traced to the treatment of the spacetime as a classical, continuum background.\footnote{We are deliberately avoiding any mention to the so-called infrared divergences in quantum field theory. If necessary, they can be prevented by, e.g., considering topologically compact spatial hypersurfaces in the considered spacetimes.} In particular, this type of spacetime description triggers the appearance of ill-defined products of field operators, which typically include the building blocks of the free field Hamiltonian in the considered background (and thus of the energy in stationary situations). 

Despite the considerable effort devoted to develop covariant renormalization techniques, even for free fields in curved spacetimes, one could be tempted to believe that, instead of recurring to those schemes for the ``substraction of infinities,'' a formalism that satisfactorily accounts for the presumable quantum nature of the spacetime would be able to prevent the occurrence of divergencies in the first place. In this sense, the role that a theory of quantum gravity might play in surpassing the limits of predictability of our current theoretical models could be twofold, actually, because it might also cure the problem of formation of spacetime singularities that is intrinsic to classical general relativity \cite{waldgr}. 

A promising candidate for the quantization of Einstein's theory is the nonperturbative and canonical formalism known as loop quantum gravity \cite{lqg}. To make direct contact with physically feasible models, the techniques developed in this formalism have been used, suitably combined with more conventional Fock quantization methods, in order to describe certain types of inhomogeneous spacetimes quantum mechanically. This procedure has been given the name of hybrid quantization, and it has been primarily applied to cosmological scenarios \cite{hybr-rev,hybr-gow1,hybr-gow2,hybr-inf1,hybr-inf2}. Essentially, this hybrid approach is based on a convenient splitting of the cosmological phase space into two sectors: a purely homogeneous one, that is represented in a quantum mechanical way by employing methods that are inspired in loop quantum gravity, and an inhomogeneous sector, for which a suitable Fock representation is adopted. In fact, the application of loop quantum gravity techniques to the quantization of homogeneous cosmologies, often known as loop quantum cosmology \cite{lqc1,lqc2,lqc3}, has been shown to lead to a quite general resolution of the cosmological singularities predicted by general relativity \cite{MMO,singh}. Remarkably, the big bang singularity is replaced with a bounce in the trajectories followed by the peaks of a wide class of quantum states in the homogeneous cosmologies studied so far in the literature (see, e.g., Refs. \cite{APS1,APS2,taveras}). 

The hybrid quantization approach extends to inhomogeneous models the expectation that, with a loop quantum cosmology representation of the homogeneous sector of the geometry, one should be able to solve (at least) the most severe singularities of a genuine cosmological nature. At the same time, this hybrid strategy gives hope for the possibility that a suitably chosen Fock representation for the inhomogeneous sector of the phase space may complete the quantum description of the system in a divergence-free way. This possibility is motivated by the existing freedom in performing canonical transformations within the entire phase space, transformations that assign different dynamical roles to the homogeneous sector of the system and to the rest of matter and gravitational degrees of freedom. Indeed, these transformations change the part of the total Hamiltonian (constraint) that, while retaining the coupling with the homogeneous sector, generates the dynamics of the inhomogeneous, fieldlike degrees of freedom. Given that each sector of the phase space is quantized in a different type of representation, it is then possible that a suitable choice of canonical transformation and Fock representation for the inhomogeneities may yield a quantum description that is free of the divergences that would otherwise appear in standard quantum field theory in curved spacetimes. In particular, this procedure would allow us to handle properly (at least certain forms of) the matter-geometry backreaction in a quantum mechanical way.

The aim of this work is to provide solid ground for our expectations by showing, in a specific cosmological system, that one can attain such a well-defined quantum hybrid description without the need of any regularization. The case that we discuss here is an inflationary homogeneous and isotropic cosmology in the presence of Dirac fermions, considered as perturbations. The hybrid quantization of this system was introduced in Ref. \cite{hybr-ferm}, allowing also for the presence of scalar and tensor perturbations of the metric and of the inflaton field, and after truncating the action at second order in all the perturbations. As far as the Dirac perturbations are concerned, the splitting of the (truncated) phase space adopted in that reference was inspired by the pioneer work in Ref. \cite{DEH} about fermions in quantum cosmology, developed in the context of quantum geometrodynamics. It was seen in Ref. \cite{hybr-ferm} that, by adopting a separation of variables between the homogeneous part of the geometry, on the one hand, and the inhomogeneities, on the other hand, in the dependence of the quantum states (separation that can be viewed as a kind of Born-Oppenheimer {\emph {ansatz}} in which the inflaton field plays the role of an internal time), it is possible to derive a quantum evolution for the fermionic perturbations that is ruled by a Schr\"odinger-like equation. Actually, the resulting dynamics is generated by the fermionic contribution to the total Hamiltonian (constraint), converting the coupling of the fermionic perturbations with the homogeneous geometry into expectation values of the corresponding geometric operators. In addition, the expectation value of this total Hamiltonian supplies information about the backreaction of the fermions (and of the rest of perturbations) on the homogeneous background. This information is given by the difference between the average of two operators on the homogeneous part of the state, difference that tells us whether such a quantum state is an exact solution of the unperturbed model or not. It was then proven in the cited work that the discussed evolution of the fermionic perturbations can be implemented unitarily in Fock space. Furthermore, explicit solutions were found by constructing an evolution operator and evolving the fermionic vacuum with it. However, it was shown that the mentioned Hamiltonian contribution of the fermionic degrees of freedom intrinsically leads to divergences (of an ultraviolet nature), with an infinite backreaction, unless one introduces a convenient regularization procedure.

In this article, we present an alternative and, at the same time, rather generic description of the system that resolves the problem of the divergences encountered in Ref. \cite{hybr-ferm}. We do so by employing in our benefit the commented freedom in adopting different dynamical splittings between the homogeneous geometric background and the fermionic perturbations, related by canonical transformations. It suffices to restrict our discussion to choices of annihilation and creationlike variables for the fermionic fields such that, when the spacetime background is considered to be classical and the fermions are treated in the context of quantum field theory in curved spacetimes, the quantum dynamics becomes unitarily implementable in Fock space (while being nontrivial according to the evolution dictated by the Dirac equation) \cite{uf-flat}. It has been shown that all such variables define unitarily equivalent Fock representations of the Dirac field, once a convention for the notions of particles and antiparticles has been set \cite{uf-flat}. In fact, the variables introduced in Ref. \cite{DEH} and then used in Ref. \cite{hybr-ferm} satisfy this unitarity condition. We characterize here the set of such annihilation and creationlike variables for which the description of the system is free of the divergences of standard quantum field theory. From a conceptual viewpoint, this result may have important implications. Moreover, it will shed light on the problem of the choice of a unique vacuum for the Dirac field in quantum cosmology (among all those available in our unitary class of Fock representations) with good physical properties.

The structure of the paper is the following. In Sec. II we summarize the description of the classical system presented in Ref. \cite{hybr-ferm}, and then introduce a more general class of annihilation and creationlike variables for the Dirac field than those adopted in that reference. With those definitions at hand, we compute the Hamiltonian that generates the associated fermionic dynamics. We start Sec. III with a brief review of the procedure to derive the corresponding Schr\"odinger equation for the fermionic degrees of freedom, after adopting a kind of Born-Oppenheimer ansatz for the physical states. In addition, we analyze the ultraviolet properties of the fermionic dynamics and deduce the conditions that the annihilation and creationlike variables must fulfil in order that their backreaction be finite. Finally, we will further impose that the Hamiltonian that drives this evolution be a well-defined operator on the fermionic vacuum and, as a consequence, a densely defined operator in Fock space. We conclude in Sec. IV with a summary of our results and a brief outlook.

\section{The classical system}

In this section we use the conventions and notation of Ref. \cite{hybr-ferm}. We refer the reader to that work for specific derivations and formulas. The starting point for the construction of the system is a Friedmann-Lema\^{\i}tre-Robertson-Walker (FLRW) spacetime with flat and compact spatial hypersurfaces (isomorphic to a three-torus, $T^3$). We employ spatial coordinates adapted to the homogeneity. The matter content is given by a homogenous scalar field subject to a potential (that, classically, would play the role of the inflaton), and a Dirac field, both of them minimally coupled. The Dirac field is treated entirely as a perturbation. Besides, we can introduce perturbations of the metric and of the scalar field, as discussed in Refs. \cite{hybr-ferm,hybr-ref,hybr-ten}. 

More specifically, we truncate our perturbed system so that its Einstein-Dirac action is at most quadratic in all the perturbations \cite{DEH,HH}. Our canonical formulation is obtained from the symplectic structure and from the Hamiltonian associated with this truncated action. Within this truncation scheme, and regardless of the consideration or not of additional perturbations, the Dirac field couples exclusively to the homogeneous tetrad that describes the FLRW sector of the cosmology, because the Dirac action is already quadratic in the fermionic field. This fact immediately implies that the fermionic degrees of freedom are gauge invariant, at the considered perturbative order. Namely, they commute under Poisson brackets with the linear perturbative (Hamiltonian and diffeomorphisms) constraints of the relativistic system. On the other hand, together with an Abelianization of these linear perturbative constraints and suitable momenta of them, it is possible to construct a completely gauge-invariant parametrization of the sector of the phase space that contains the physical information about the metric and scalar field perturbations, as explained in Refs. \cite{hybr-ref,hybr-ten}. In particular, this information can be encoded in a set of variables that consists of the well-known tensor and Mukhanov-Sasaki gauge invariants \cite{bardeen,gaugeinvariant,sasa,kodasasa,mukhanov}. To arrive at this description of the perturbations, one introduces linear transformations on the original perturbative variables, transformations which depend on the homogeneous sector of the phase space. It is then possible to complete the change of variables to include this homogeneous sector as well and obtain a canonical set for the entire system, again at the considered truncation order in perturbations. As a result, the new canonical variables for the homogeneous degrees of freedom acquire a (spatially integrated) correction which is quadratic in the metric and inflaton perturbations. In the case of the Dirac field, given its consideration as a perturbation and the fact that its contribution to the action is quadratic, one finds within our truncation scheme that the expression of the Dirac Hamiltonian in terms of the new homogeneous tetrads amounts just to a minimal coupling of the fermions directly with such new variables.

To exploit the spatial symmetries associated with the homogeneous foliation of the unperturbed sector of our model, that is, the FLRW cosmology, it is most convenient to expand the perturbations in spatial modes (of the Laplace-Beltrami or Dirac operators) on $T^3$. For the fermionic content, in particular, each of the two chiral components of the Dirac field may be expanded in a complete set of eigenspinors of the Dirac operator on $T^3$, after imposing the time gauge on the homogeneous tetrads (e.g., by considering a diagonal gauge) \cite{DEH}. The spectrum of that operator is discrete and characterized by eigenvalues $\pm\omega_{k}=\pm2\pi|\vec{k}+\vec{\tau}|/l_0$, where $l_0$ is the compactification length of the tori, $\vec{k}\in\mathbb{Z}^3$, and $2\vec{\tau}$ can be any of the constant vectors that form the standard orthonormal basis of the lattice $\mathbb{Z}^3$ and that characterize each of the eight possible spin structures on $T^3$ \cite{SGeom,Dtorus}. Since $\omega_k$ grows like $|\vec{k}|$ when this quantity tends to infinity, the density of states with eigenvalues in an interval $(\omega_k,\omega_k+\Delta\omega_k]$ grows asymptotically as $\omega_k^2\Delta\omega_k$ multiplied by a constant. Then, let us consider the Dirac field multiplied by the rescaling factor $e^{3\alpha/2}$, where $\alpha$ is, up to an additive constant, the logarithm of the scale factor of the FLRW cosmology, once we have corrected it with quadratic contributions of the perturbations as we have commented above. In the expansion of the left-handed component of such rescaled Dirac field, we call $m_{\vec{k}}$ and $\bar{r}_{\vec{k}}$, up to a multiplicative constant $[4\pi/(3l_0)]^{-3/4}$, the time-dependent coefficients of the eigenspinors of the Dirac operator on $T^3$ with respective eigenvalues $\omega_k$ and $-\omega_k$. Similarly, $\bar{s}_{\vec{k}}$ and $t_{\vec{k}}$ respectively denote the coefficients, up to the mentioned constant factor, of the complex conjugates of the eigenspinors with eigenvalues $\omega_k$ and $-\omega_k$ in the expansion of the right-handed component of the rescaled Dirac field. All of these eigenspinor coefficients are taken as Grassmann variables \cite{Berezin}, in order to capture the anticommuting nature of the field. Besides, each of them forms a canonical pair with its complex conjugate, with a Dirac bracket (obtained after eliminating second-class constraints that relate the Dirac field with its momentum) equal to $-i$, and vanishing anticommutation relations with the rest of coefficients. Introducing these mode decompositions in the action, one obtains the fermionic contribution to the total Hamiltonian. This contribution is quadratic in the fermionic variables, and is given by a sum over all modes, which decouple from each other. It comes multiplied by the  homogeneous lapse function $N_0$, so we call it $N_{0}H_D$. As expected, this fermionic term adds to the zero mode of the Hamiltonian constraint, which is therefore the only constraint affected.

\subsection{Instantaneous diagonalization of the Dirac Hamiltonian}

As commented in the introduction, and partially motivated by the work of D'Eath and Halliwell \cite{DEH}, the following annihilation and creationlike variables were chosen in Ref. \cite{hybr-ferm} for the description of the fermionic degrees of freedom:
\begin{eqnarray}\label{DH}
\breve{a}_{\vec{k}}^{(x,y)}&=& \sqrt{ \frac{\xi_k -\omega_k}{2\xi_k} }\, x_{\vec{k}}+\sqrt{ \frac{\xi_k +\omega_k}{2\xi_k}}\, {\bar y}_{-\vec{k}-2\vec{\tau}}, \nonumber\\
{\bar {\breve{b}}}_{\vec{k}}^{(x,y)}&=& \sqrt{ \frac{\xi_k +\omega_k}{2\xi_k}}\, x_{\vec{k}}-\sqrt{ \frac{\xi_k -\omega_k}{2\xi_k} } \,{\bar y}_{-\vec{k}-2\vec{\tau}},	
\end{eqnarray}
where $(x_{\vec{k}},y_{\vec{k}})$ is any of the ordered pairs $(m_{\vec{k}},s_{\vec{k}})$ or $(t_{\vec{k}},r_{\vec{k}})$, and $\breve{a}_{\vec{k}}^{(x,y)}$ and ${\bar {\breve{b}}}_{\vec{k}}^{(x,y)}$ correspond to annihilationlike variables for particles and creationlike variables for antiparticles, respectively. Besides, an overbar denotes complex conjugation, and we have defined
\begin{equation}
\label{xi}
\xi_k=\sqrt{\omega_k^2+\tilde{M}^2 e^{2\alpha}},
\end{equation}
where $\tilde{M}=2 M \sqrt{\pi/(3l_0^3)}$ is the mass $M$ of the Dirac field up to a multiplicative constant \cite{hybr-ferm}. Notice that then the square roots appearing in Eq. \eqref{DH} are always well defined and real. The variables \eqref{DH} are distinguished (apart from irrelevant redefinitions among degenerate modes) by the fact that they diagonalize $H_D$ (if one ignores the $\vec{k}=\vec{\tau}$ mode\footnote{This particular contribution to $H_D$ is only present when a trivial spin structure is chosen on $T^3$, and it corresponds to $\omega_k=0$. We will safely ignore it throughout this work since, owing to the compactness of the spatial sections, it can be isolated from the rest of contributions and be handled without producing infrared divergences.}). This diagonalization means that no term containing creation or annihilation of particle-antiparticle pairs appears in the resulting expression of $H_D$. More specifically, if we call $H_{\vec{k}}$, with $\vec{k}\neq \vec{\tau}$, each of the terms in the sum that forms $H_D$, we get
\begin{eqnarray}
\label{HkDEH}
{H}_{\vec{k}} &=& \frac{e^{-{ \alpha}}} {2} \sum_{(x,y)} \bigg[ \xi_k \bigg( {\bar{ \breve{a}}}_{\vec{k}}^{(x,y)}\breve{a}_{\vec{k}}^{(x,y)}-  \breve{a}_{\vec{k}}^{(x,y)}{\bar{\breve{ a}}}_{\vec{k}}^{(x,y)} +{\bar{\breve{ b}}}_{\vec{k}}^{(x,y)}\breve{b}_{\vec{k}}^{(x,y)}  - \breve{b}_{\vec{k}}^{(x,y)}{\bar{\breve{ b}}}_{\vec{k}}^{(x,y)} \bigg) \bigg],
\end{eqnarray} 
where the sum over $(x,y)$ is over the pairs $(m,s)$ and $(t,r)$. Although this diagonalization might seem appealing, it turns out that the introduction of these annihilation and creationlike variables gives rise to the appearance of an additional, nondiagonal, quadratic contribution to the fermionic part of the Hamiltonian of the system. This is due to the fact that the definition \eqref{DH} is a background-dependent linear transformation of the fermionic mode coefficients, inasmuch as it involves the homogeneous variable $\alpha$. In fact, it is not hard to see that the transformation is canonical when restricted to the fermionic sector of the phase space. However if, adopting the strategy of the hybrid approach, one wants a transformation that respects the canonical symplectic structure of the entire set of degrees of freedom at the considered order of truncation, then the momentum of $\alpha$ must be modified with the addition of a factor that is quadratic in the fermionic perturbations, according to our previous comments. If we call $\breve\pi_{\alpha}$ this new canonical momentum, the expression of the total Hamiltonian in terms of the new canonical variables is functionally the same as in terms of the old homogeneous ones, but with an additional sum over $\vec{k}$ of the following contributions \cite{hybr-ferm}:

\begin{eqnarray}
\label{intDEH}
-i N_0 \sum_{(x,y)} 
\frac{\tilde{M}\omega_k }{{2 \xi}_k ^2} e^{-{2 \alpha}}{\breve \pi}_{\alpha}  \bigg( \breve{a}_{\vec{k}}^{(x,y)} \breve{b}_{\vec{k}}^{(x,y)} + {\bar {\breve{a}}}_{\vec{k}}^{(x,y)} {\bar {\breve{b}}}_{\vec{k}}^{(x,y)}\bigg).
\end{eqnarray} 
The coefficient of each of these ``interaction" terms, that produce the creation and annihilation of pairs, decays asymptotically as $\omega_k^{-1}$. As shown in Ref. \cite{hybr-ferm}, this asymptotic behavior is transmitted to the quantum theory [at least in regimes where the state for the homogeneous geometry experiences (almost) no transition mediated by the total Hamiltonian, so that the geometric information can be encoded in expectation values on this state]. This asymptotic behavior, together with the specific dependence on $\omega_k$, $\alpha$, and $\tilde{M}$ of the part of $H_{\vec{k}}$ which is asymptotically dominant, is what at the end of the day guarantees that the fermionic quantum dynamics can be implemented unitarily in Fock space. Nonetheless, the fact that the discussed interaction terms decay as $\omega_k^{-1}$ in the ultraviolet regime, and not faster, is precisely what leads to a possibly divergent backreaction on the state of the homogenous geometry. Indeed, such backreaction was seen to be a sum over $\vec{k}$ of terms of dominant order equal to $\omega_k^{-3}$, which is not absolutely convergent, given the quadratic growth of the density of states (see e.g. Refs. \cite{torus,torus1} for additional details concerning the convergence of mode-dependent series in $T^3$).

\subsection{Alternative choices of fermionic variables}

In order to explore whether other choices of fermionic variables may elude the appearance of divergences in the quantum field theory treatment, in this section we will consider a rather generic family of alternative definitions of annihilation and creationlike variables for the Dirac field. For this purpose, we will exploit the freedom to perform linear canonical transformations of the fermionic variables that depend on the homogeneous background geometry. In doing it, we are contemplating the possibility of considering different dynamical splittings between the background geometry and the genuine fermionic degrees of freedom. This possibility comes naturally on stage when one aims at constucting a quantum mechanical description of the system as a whole, following a hybrid scheme in which the homogeneous sector of the phase space is represented in a fundamentally different manner.

Obviously, when one adopts this perspective, the choice of fermionic variables is affected by a vast ambiguity. This ambiguity can be viewed as twofold. On the one hand, there are certainly many ways of redefining the dynamical behavior of the fermionic excitations (and, correspondingly, of the cosmological variables) by plugging different dependencies on $\alpha$ and its momentum $\pi_\alpha$ in the linear canonical transformations that define the fermionic variables. On the other hand, even after a dynamical splitting has been set, choices of fermionic annihilation and creationlike variables related by constant transformations can give rise to different, and in many cases inequivalent, Fock representations, each with its associated vacuum. Actually, both types of ambiguities can be analyzed simultaneously, restricting to choices that respect the dynamical decoupling between modes, by introducing generic annihilation and creationlike variables of the form
\begin{eqnarray}
\label{ab}
a_{\vec{k}}^{(x,y)}&=& f_1^{\vec{k},(x,y)}(\alpha,\pi_{\alpha})\, x_{\vec{k}}+f_2^{\vec{k},(x,y)}(\alpha,\pi_{\alpha})\, {\bar y}_{-\vec{k}-2\vec{\tau}}, \nonumber\\
{\bar b}_{\vec{k}}^{(x,y)}&=& g_1^{\vec{k},(x,y)}(\alpha,\pi_{\alpha})\, x_{\vec{k}}+g_2^{\vec{k},(x,y)}(\alpha,\pi_{\alpha}) \,{\bar y}_{-\vec{k}-2\vec{\tau}},	
\end{eqnarray}
where, to satisfy the standard canonical anticommutation relations, one must have \cite{hybr-ferm}
\begin{eqnarray}
\label{gs}
g^{\vec{k},(x,y)}_{1} &=& e^{iJ_{\vec{k}}^{(x,y)}}{\bar f}^{\vec{k},(x,y)}_{2},\qquad g^{\vec{k},(x,y)}_{2} = - e^{iJ_{\vec{k}}^{(x,y)}} {\bar f}^{\vec{k},(x,y)}_{1} ,\\
\label{ftwo}
f^{\vec{k},(x,y)}_{2}&=& e^{iF^{\vec{k},(x,y)}_{2}} \sqrt{1- \left\vert f^{\vec{k},(x,y)}_{1} \right\vert^2},
\end{eqnarray}
with $J_{\vec{k}}^{(x,y)}$ and $F^{\vec{k},(x,y)}_{2}$ being some (possibly background-dependent) phases. Clearly, the choice \eqref{DH} is one of these many different sets of annihilation and creationlike variables.

Despite all the freedom allowed in the definitions \eqref{ab}, one can restrict the selection of annihilation and creationlike variables to a single privileged family of unitarily equivalent choices by imposing some physically desirable properties. In this sense, a satisfactory criterion is the imposition that the dynamics of the annihilation and creationlike variables can be implemented as unitary transformations in Fock space (for dynamics that are not rendered trivial with respect to the evolution dictated by the Dirac equation and when the Dirac field is treated as a test field propagating on the FLRW cosmology). This condition, together with the invariance of the vacuum under the continuous isometries of the toroidal sections of the homogeneous cosmology, and a standard convention for the notions of particles and antiparticles, indeed leads to a family of unitarily equivalent Fock representations \cite{uf-flat}. Actually, the set of annihilation and creationlike variables defined in Eq. \eqref{DH} belongs to this privileged family (this was precisely the motivation to adopt that set in Ref. \cite{hybr-ferm}). Going beyond this particular choice, which we recall diagonalizes $H_D$, it turns out that the family of fermionic variables \eqref{ab}-\eqref{ftwo} that satisfies the explained selection criterion is totally specified by the following asymtpotic behavior in the limit of large $\omega_k$:
\begin{eqnarray}\label{unitf1}
f^{\vec{k},(x,y)}_{1}&=&\sqrt{ \frac{\xi_k -\omega_k}{2\xi_k} }+\frac{\tilde{M}e^{\alpha}}{2\omega_{k}}\left[e^{iF^{\vec{k},(x,y)}_{2}}-1\right]+\theta_{\vec{k}}^{(x,y)}\quad {\mathrm{with}}\quad
\sum_{\vec{k}} \left\vert\theta_{\vec{k}}^{(x,y)}\right\vert^{2}<\infty.
\end{eqnarray}
More specifically, the sequence $\{\theta_{\vec{k}}^{(x,y)}\}_{\vec{k}\in\mathbb{Z}^3}$ must contain an infinite subsequence that is $o(\omega_k^{-1})$, where the symbol $o(.)$ means asymptotically negligible with respect to its argument. In fact, given the asymptotic behavior of the Dirac eigenvalues and of their density of states, and hence the generic nonsummability of the sequence $\omega_k^{-3}$ over all $\vec{k}\in\mathbb{Z}^{3}$, it is not hard to convince oneself that $\theta_{\vec{k}}^{(x,y)}$ must have the following asymptotic behavior. For a nonempty and infinite subset $\tilde{\mathbb{Z}}^3\subset\mathbb{Z}^3$, the functions $\theta_{\vec{k}}^{(x,y)}$ with $\vec{k}\in\tilde{\mathbb{Z}}^3$ must be $o(\omega_k^{-3/2})$. In addition to this, there might exist a complementary subset $\mathbb{Z}^3_{\uparrow}$ of infinite cardinal such that the sequence $\{\theta_{\vec{k}}^{(x,y)}\}_{\vec{k}\in\mathbb{Z}^3_{\uparrow}}$, while being square summable, is of asymptotic order $\omega_k^{-3/2}$ or higher. 

On the other hand, let us recall that both $F_{2}^{\vec{k},(x,y)}$ and $\theta_{\vec{k}}^{(x,y)}$ may be functions of the homogeneous FLRW variables $(\alpha,\pi_\alpha)$. For the sake of concreteness in our analysis and adopting in the following the notation $\{h_l\}=\{f_l,g_l\}$, with $l=1,2$, for any of the functions that determine the fermionic variables, we restrict ourselves to functional dependencies such that
\begin{equation}\label{dcond}
\partial^{n}_{\alpha}h_{l}^{\vec{k},(x,y)}=\mathcal{O}(h_{l}^{\vec{k},(x,y)}),\qquad \partial^{n}_{\pi_\alpha}h_{l}^{\vec{k},(x,y)}=\mathcal{O}(h_{l}^{\vec{k},(x,y)}),
\end{equation}
for integers $n$ at least up to 3 (and where the derivatives act order by order in the asymptotic expansion for large $\omega_k$, at least for the relevant orders in our discussion). Here,  a contribution is $\mathcal{O}(.)$ when it is of the asymptotic order of the corresponding argument (or smaller). Our restriction excludes, in particular, the possibility of absorbing in the phases of $h^{\vec{k},(x,y)}_{1}$ and $h^{\vec{k},(x,y)}_{2}$ any of the dominant oscillations in conformal time that the Dirac field displays in the limit of large $\omega_k$ when it is treated as a test field obeying the Dirac equation in a classical FLRW cosmology.

Similar to the situation found in the previous subsection, the family of annihilation and creationlike variables defined by Eqs. \eqref{ab}-\eqref{ftwo}, together with condition \eqref{unitf1}, is obtained by means of an $(\alpha,\pi_{\alpha})$-dependent transformation that is canonical within the fermionic sector of the phase space. In order to be canonical in the entire truncated system, as desired e.g. in the hybrid quantization strategy, the geometric variables $(\alpha,\pi_{\alpha})$ of the homogeneous sector must be replaced with a new, corrected, canonical pair $(\tilde{\alpha},\tilde{\pi}_{\alpha})$. Concretely, the corrections $\Delta\tilde\alpha=\tilde{\alpha}-\alpha$ and $\Delta\tilde{\pi}_{\alpha}=\tilde{\pi}_{\alpha}-\pi_{\alpha}$ that determine these new variables are quadratic in the fermionic perturbations, and are given by \cite{hybr-ferm} 
\begin{eqnarray}
\label{newscale}
\Delta\tilde\alpha&=&\frac{i}{2}\sum_{\vec{k},(x,y)}[
(\partial_{\pi_{\alpha}}x_{\vec{k}})  {\bar x}_{\vec{k}}+(\partial_{\pi_{\alpha}}{\bar x}_{\vec{k}})  x_{\vec{k}}+(\partial_{\pi_{\alpha}}y_{\vec{k}})  {\bar y}_{\vec{k}}+(\partial_{\pi_{\alpha}}{\bar y}_{\vec{k}})  y_{\vec{k}} ],
\\
\label{newpi}
\Delta\tilde{\pi}_{\alpha}&=&-\frac{i}{2}\sum_{\vec{k},(x,y)}[
(\partial_{\alpha}x_{\vec{k}})  {\bar x}_{\vec{k}}+(\partial_{\alpha}{\bar x}_{\vec{k}})  x_{\vec{k}}+(\partial_{\alpha}y_{\vec{k}})  {\bar y}_{\vec{k}}+(\partial_{\alpha}{\bar y}_{\vec{k}})  y_{\vec{k}} ].
\end{eqnarray} 
Taking into account the quadratic order of our perturbative truncation, one then concludes that the expression of the total Hamiltonian of the cosmological system in terms of these new variables can be obtained by directly substituting the new pair $(\tilde{\alpha},\tilde{\pi}_{\alpha})$ in its functional dependence on $(\alpha,\pi_{\alpha})$, and replacing the Dirac Hamiltonian $N_0H_D$ with
\begin{equation}\label{tildeH}
N_0{\tilde H}_D=N_0\left[H_D+e^{-3{\tilde \alpha}}{\tilde \pi}_{\alpha}\Delta\tilde\pi_{\alpha}-8\pi e^{3\tilde\alpha} V(\phi)\Delta\tilde\alpha\right].
\end{equation}
Here, $V(\phi)$ is (up to a multiplicative constant \cite{hybr-ferm}) the potential of the homogeneous inflaton field $\phi$, and all the dependence of $H_D$, $\Delta\tilde\alpha$, and $\Delta\tilde{\pi}_{\alpha}$ on the homogeneous pair $(\alpha,\pi_{\alpha})$ must again be evaluated at $(\tilde{\alpha},\tilde{\pi}_{\alpha})$. In order to arrive at this corrected fermionic Hamiltonian, a well-controlled redefinition of the homogeneous lapse function must be performed, adding to it a sum over modes of certain terms that are quadratic in the fermionic perturbations \cite{hybr-ferm}.

Let us notice that, in terms of the family of annihilation and creationlike variables \eqref{ab}-\eqref{unitf1} that we are considering, the Dirac contribution $H_D$ to the Hamiltonian does no longer, in general, display a diagonal form as it did before [see Eq. \eqref{HkDEH}]. In fact, one may obtain the new expression of $H_D$ by inserting in Eq. \eqref{HkDEH} the Bogoliubov transformation
\begin{eqnarray}\label{Bog1}
\breve{a}_{\vec{k}}^{(x,y)}&=& \kappa^{(x,y)}_{\vec{k}}a_{\vec{k}}^{(x,y)}+\lambda^{(x,y)}_{\vec{k}}\bar{b}_{\vec{k}}^{(x,y)}, \nonumber\\\label{Bog2}
\bar{\breve{b}}_{\vec{k}}^{(x,y)}&=&e^{-iJ_{\vec{k}}^{(x,y)}}\left[\bar{\kappa}^{(x,y)}_{\vec{k}}\bar{b}_{\vec{k}}^{(x,y)} -\bar{\lambda}^{(x,y)}_{\vec{k}}a_{\vec{k}}^{(x,y)}\right],
\end{eqnarray}
that relates the old variables $\{\breve{a}_{\vec{k}}^{(x,y)},\bar{\breve{b}}_{\vec{k}}^{(x,y)}\}$ employed in Refs. \cite{hybr-ferm,DEH} with the more general family considered here. It is not hard to check that relations \eqref{gs},\eqref{ftwo} guarantee that this is indeed a Bogoliubov transformation in the fermionic phase space, so that in particular we have $|\kappa^{(x,y)}_{\vec{k}}|^2+|\lambda^{(x,y)}_{\vec{k}}|^2=1$. A straightforward computation then shows that
\begin{align}
{H}_{\vec{k}} &= \frac{e^{-{ \tilde\alpha}}} {2} \sum_{(x,y)} \bigg[ \tilde\xi_k \bigg(1-2|\lambda^{(x,y)}_{\vec{k}}|^2\bigg)\bigg( {\bar{ {a}}}_{\vec{k}}^{(x,y)}{a}_{\vec{k}}^{(x,y)}-  {a}_{\vec{k}}^{(x,y)}{\bar{{ a}}}_{\vec{k}}^{(x,y)} +{\bar{{ b}}}_{\vec{k}}^{(x,y)}{b}_{\vec{k}}^{(x,y)}  - {b}_{\vec{k}}^{(x,y)}{\bar{{ b}}}_{\vec{k}}^{(x,y)} \bigg)\nonumber \\&-4\tilde\xi_k \bigg( {\kappa}^{(x,y)}_{\vec{k}}\bar{\lambda}^{(x,y)}_{\vec{k}}{a}_{\vec{k}}^{(x,y)} {b}_{\vec{k}}^{(x,y)} -\bar{\kappa}^{(x,y)}_{\vec{k}}{\lambda}^{(x,y)}_{\vec{k}} {\bar {{a}}}_{\vec{k}}^{(x,y)} {\bar {{b}}}_{\vec{k}}^{(x,y)}\bigg) \bigg],
\end{align}
where ${\tilde \xi}_k$ stands for the result of replacing $\alpha$ directly with $\tilde{\alpha}$ in the definition \eqref{xi} of $\xi_k$. Besides, we recall that $H_D$ is the sum over all modes $\vec{k}\neq\vec{\tau}$ of the corresponding Hamiltonian term ${H}_{\vec{k}}$.

Apart from the mentioned contributions to $H_D$, interaction terms that cause the creation and annihilation of pairs in all modes arise again from the corrections that are proportional to $\Delta\tilde\alpha$ and $\Delta\tilde\pi_{\alpha}$ in the expression \eqref{tildeH} of the fermionic Hamiltonian $N_0\tilde{H}_D$. All those terms can be computed using Eqs. \eqref{newscale} and \eqref{newpi} after imposing the asymptotic relations \eqref{unitf1}. Then, one can regard the resulting fermionic Hamiltonian as a sum over all $\vec{k}\in\mathbb{Z}^3$ of some functions $N_0\tilde{H}_{\vec{k}}$ that possess a quite specific asymptotic behavior. One obtains
\begin{align}\label{tildeHk}
\tilde{H}_{\vec{k}}&= \sum_{(x,y)} \bigg[\bigg( \frac{e^{-{ \tilde\alpha}}} {2}\tilde\xi_k+h_D^{\vec{k}}\bigg) \bigg( {\bar{ {a}}}_{\vec{k}}^{(x,y)}{a}_{\vec{k}}^{(x,y)}-  {a}_{\vec{k}}^{(x,y)}{\bar{{ a}}}_{\vec{k}}^{(x,y)} +{\bar{{ b}}}_{\vec{k}}^{(x,y)}{b}_{\vec{k}}^{(x,y)}  - {b}_{\vec{k}}^{(x,y)}{\bar{{ b}}}_{\vec{k}}^{(x,y)} \bigg)+h_{J}^{\vec{k}}\bigg({\bar{{ b}}}_{\vec{k}}^{(x,y)}{b}_{\vec{k}}^{(x,y)}  - {b}_{\vec{k}}^{(x,y)}{\bar{{ b}}}_{\vec{k}}^{(x,y)}\bigg)\nonumber \\& + e^{i(J_{\vec{k}}^{(x,y)}-F_{2}^{\vec{k},(x,y)})}e^{-\tilde{\alpha}} \bigg(2\omega_k \bar{\theta}_{\vec{k}}^{(x,y)}+\bar{h}_{I}^{\vec{k}}\bigg) {a}_{\vec{k}}^{(x,y)} {b}_{\vec{k}}^{(x,y)} -e^{-i(J_{\vec{k}}^{(x,y)}-F_{2}^{\vec{k},(x,y)})} e^{-\tilde{\alpha}} \bigg(2\omega_k {\theta}_{\vec{k}}^{(x,y)}+{h}_{I}^{\vec{k}}\bigg) {\bar {{a}}}_{\vec{k}}^{(x,y)} {\bar {{b}}}_{\vec{k}}^{(x,y)} \bigg],
\end{align}
where we have defined
\begin{equation}\label{hj}
h_{J}^{\vec{k}}=-4\pi e^{3\tilde{\alpha}}V(\phi)\partial_{ \tilde{\pi}_{\alpha}}J_{\vec{k}}^{(x,y)}(\tilde{\alpha},\tilde{\pi}_{\alpha})-\frac{1}{2}e^{-3\tilde{\alpha}}\tilde{\pi}_{\alpha}\partial_{\tilde{\alpha}}J_{\vec{k}}^{(x,y)}(\tilde{\alpha},\tilde{\pi}_{\alpha}).
\end{equation}
To avoid complicating the notation in excess, we denote the partial derivatives with respect to the homogeneous geometry evaluated at $(\tilde{\alpha},\tilde{\pi}_{\alpha})$ directly by $\partial_{\tilde{\alpha}}$ and $\partial_{ \tilde{\pi}_{\alpha}}$.
Besides, $h_D^{\vec{k}}$ is a real function that, in the asymptotic regime of large $\omega_k$, is given by
\begin{equation}\label{hd}
h_D^{\vec{k}}=4\pi e^{4\tilde{\alpha}}V(\phi)\partial_{\tilde{\pi}_{\alpha}}F_{2}^{\vec{k},(x,y)}(\tilde{\alpha},\tilde{\pi}_{\alpha})+\frac{1}{2}e^{-2\tilde{\alpha}}\tilde{\pi}_{\alpha}\partial_{\tilde{\alpha}}F_{2}^{\vec{k},(x,y)}(\tilde{\alpha},\tilde{\pi}_{\alpha})+\mathcal{O}\left(\mathrm{Max}\big[\omega_k^{-2},(\theta_{\vec{k}}^{(x,y)})^2\big]\right).
\end{equation}
In this asymptotic regime, we also have for $\vec{k}\in\mathbb{Z}^{3}_{\uparrow}$,
\begin{equation}
\label{hi1}
h_I^{\vec{k}}=\mathcal{O}\left(\mathrm{Max}\big[\omega_k^{-1},\theta_{\vec{k}}^{(x,y)},\omega_k(\theta_{\vec{k}}^{(x,y)})^3\big]\right),
\end{equation}
while, for $\vec{k}\in\tilde{\mathbb{Z}}^{3}$,
\begin{eqnarray}\label{hi2}
h_I^{\vec{k}}&=&ie^{-2\tilde{\alpha}}\tilde{\pi}_{\alpha}\bigg[\frac{\tilde{M}e^{\tilde{\alpha}}}{2\omega_k}e^{iF^{\vec{k},(x,y)}_2}+\partial_{\tilde{\alpha}}\theta^{(x,y)}_{\vec{k}}(\tilde{\alpha},\tilde{\pi}_{\alpha})-i{\theta}^{(x,y)}_{\vec{k}}\partial_{\tilde{\alpha}}F_{2}^{\vec{k},(x,y)}(\tilde{\alpha},\tilde{\pi}_{\alpha})\bigg]\nonumber\\
&+&8\pi ie^{4\tilde{\alpha}}V(\phi)\bigg[\partial_{\tilde{\pi}_{\alpha}}\theta^{(x,y)}_{\vec{k}}(\tilde{\alpha},\tilde{\pi}_{\alpha})-i{\theta}^{(x,y)}_{\vec{k}}\partial_{\tilde{\pi}_{\alpha}}F_{2}^{\vec{k},(x,y)}(\tilde{\alpha},\tilde{\pi}_{\alpha})\bigg]+\mathcal{O}(\omega_k^{-2}). 
\end{eqnarray}
The function $\mathrm{Max}[.,.]$ picks out the argument of dominant asymptotic order. To arrive at these expressions, we have made a convenient use of condition \eqref{dcond}. Given the standard convention for the assignation of particles and antiparticles, this is the only relevant restriction that we impose on the family of annihilation and creationlike variables, apart from the physically appealing requirement of a quantum dynamics that is compatible with the symmetries of the homogeneous cosmology and is unitarily implementable, in the context of quantum field theory in curved spacetimes.

\section{Backreaction term in the Hamiltonian}

The asymptotic characterization that we have carried out of the fermionic part $\tilde{H}_D$ in the zero mode of the Hamiltonian constraint allows for a rather general passage to the quantum theory, without the need to specify a particular choice of fermionic annihilation and creationlike variables [among those allowed by Eqs. \eqref{unitf1} and \eqref{dcond}]. With that freedom in mind, we now briefly summarize the hybrid quantization of the system and display the equations that result for the fermionic perturbations when one adopts a kind of Born-Oppenheimer ansatz for the quantum states. We recall that the phase space of the system has been split into the following sectors. First of all, there is the homogeneous background, with canonical variables that, after being perturbatively corrected, describe the homogeneous FLRW geometry and the homogeneous inflaton. Secondly, we have the information about the scalar and tensor perturbations, encoded in the tensor and Mukhanov-Sasaki gauge invariants, as well as in the linear perturbative constraints of the system, together with their canonical momenta. Finally, the fermionic degrees of freedom are characterized by variables of the form \eqref{ab}-\eqref{ftwo} subject to the conditions \eqref{unitf1} [and \eqref{dcond}]. All of these sectors are jointly subject to the zero mode of the Hamiltonian constraint, formed from the constraint of the unperturbed inflationary model (but evaluated now in the new, corrected, background variables) by adding to it terms that are quadratic in the gauge-invariant perturbations. In particular, $\tilde{H}_D$ provides the fermionic contribution to this global constraint. In the hybrid approach, one then adopts some suitably chosen quantum representations for each of the different sectors, each of them with its corresponding Hilbert or Fock space, and introduces some well-defined operator(s) on the resulting tensor product space to represent the constraint(s), imposed quantum mechanically. This is highly nontrivial, given the fact that the zero mode of the Hamiltonian constraint mixes the homogeneous sector, which is provided with a quantum gravity-inspired representation, with all the rest.

In this work, we do not worry about the specific details of the representation chosen for the tensor and Mukhanov-Sasaki perturbations, or about their associated part of the zero mode of the Hamiltonian constraint. It suffices to say that they are described with a suitable Fock representation (for additional details, see e.g. Refs. \cite{hybr-ferm,hybr-ref,hybr-inf3}). As for the Abelianized, linear perturbative constraints, their imposition can be made straightforward, since they are part of the constructed set of canonical variables. They just restrict the quantum states not to depend on their canonical momenta, which are purely gauge degrees of freedom. The remaining sectors that are relevant for our study are then the homogeneous background and the fermionic perturbations. For the former, we select a loop quantum cosmology-inspired representation \cite{MMO,APS2}. In short, this means that, instead of working with the canonical pair $(\tilde{\alpha},\tilde{\pi}_{\alpha})$, one performs a canonical transformation to obtain a new pair that describes (up to corrections that are quadratic in perturbations) the physical volume of the universe $V$ and its canonical momentum. This latter variable contains, in turn, the information about the Ashtekar-Barbero connection for the homogeneous sector. The volume variable and the complex exponentiation of its momentum are then the functions of the homogeneous geometry that are represented quantum mechanically, adopting what is known as a polymeric representation. It is common to construct it on a Hilbert space formed from eigenstates of the volume, with the discrete inner product \cite{lqc3}. We denote this polymeric Hilbert space as $\mathcal{H}_\text{kin}^\text{grav}$. On the other hand, for the inflaton field $\phi$ and its momentum we choose a standard Schr\"odinger representation, with Hilbert space given by the space of square integrable functions of the inflaton, $L^2(\mathbb{R},d\phi)$. And for the fermionic perturbations we consider the Fock representation associated with any choice of annihilation and creationlike variables within the family defined by Eqs. \eqref{ab}-\eqref{dcond}.  We call $\mathcal{F}_D$ the corresponding Fock space. Besides ${\hat a}_{\vec{k}}^{(x,y)}$ and ${\hat b}_{\vec{k}}^{(x,y)\dagger}$ respectively denote the annihilation operators of particle excitations and the creation operators of antiparticle excitations, with their adjoints acting reversely. Let us recall that all the possible Fock representations chosen in this way are unitarily equivalent. However, as we have seen in the previous section, the fermionic Hamiltonian, and in particular its asymptotic tail in the mode decomposition with respect to the eigenspinors of the Dirac operator in $T^3$, can experience significant changes when choosing different annihilation and creationlike variables in the considered family. It is this freedom what we now exploit in order to see whether we can avoid the appearance of ultraviolet divergences in the quantum theory.

With the representation space fixed as the tensor product of all the mentioned spaces, the construction of an operator for the zero mode of the Hamiltonian constraint involves some additional choices. For the representation of the nonpolynomic functions of the homogeneous variables that appear in the different contributions to the constraint, we refer the reader to the prescriptions listed in Refs. \cite{hybr-ferm,hybr-inf3}. It suffices to say here that it is possible to define them in such a way that the action of the constraint divides the space $\mathcal{H}_\text{kin}^\text{grav}$ into separable sectors (called superselection sectors) which provide a strictly positive lower bound for the homogeneous volume $V$ \cite{MMO}. On the other hand, we impose normal ordering for the annihilation and creation operators that represent the Fock quantized perturbations.

\subsection{Schr\"odinger and Heisenberg equations}

In order to find solutions to the zero mode of the Hamiltonian constraint, namely states that are annihilated by its (adjoint) action, we follow the strategy of Refs. \cite{hybr-ferm,hybr-ref,hybr-inf3} and adopt a convenient ansatz  as follows. We consider states with a wave function in which the dependence on the homogeneous geometry and on each of the perturbative sectors can be factorized in a different term. On the other hand, all of these factors, that can be regarded as wave functions for each of the corresponding sectors, are allowed to depend on the homogeneous inflaton, $\phi$, which then plays the role of an internal time for the total system. We generically call $\Gamma(V,\phi)$ the part of the wave function that contains the information about the homogeneous geometry, while $\psi_D(\mathcal{N}_D,\phi)$ denotes the part with dependence on the fermionic degrees of freedom. The abstract notation $\mathcal{N}_D$ refers to the occupation numbers of all the fermionic particles and antiparticles. Moreover, as an ingredient of our ansatz, we restrict our considerations to normalized states $\Gamma$ in $\mathcal{H}_\text{kin}^\text{grav}$ with a unitary evolution in $\phi$, which furthermore is generated by a positive operator $\hat{\tilde{\mathcal H}}_0$,
\begin{align}
-i\partial_{\phi}\Gamma(V,\phi)=\hat{\tilde{\mathcal H}}_0\Gamma(V,\phi).
\end{align}
Besides, the above generator is chosen so that the action of $(\hat{\tilde{\mathcal H}}_0)^2+\partial_{\phi}^2$ on $\Gamma$ differs from the corresponding action of the constraint of the unperturbed FLRW cosmology at most in a quadratic contribution of the perturbations. 

With this ansatz for the states, we impose the Hamiltonian constraint (conveniently densitized in the homogeneous volume). Then, if in the state $\Gamma$ we can ignore any transition in the homogeneous geometry mediated by the action of our quantum Hamiltonian constraint, and the contribution of the perturbations to the momentum of the inflaton is negligible with respect to that of $\Gamma$ (estimated as the expectation value of $\hat{\tilde{\mathcal H}}_0$), we arrive at a collection of Schr\"odinger-like equations, with respect to $\phi$, one for each of the partial wave functions of the system on the different perturbative sectors. For details about the calculations and involved approximations, we refer the reader to Refs. \cite{hybr-ferm,hybr-ref}. Here we are interested in the equation that rules the evolution of the fermionic wave function $\psi_D$ with respect to $\phi$. This equation was deduced in Ref. \cite{hybr-ferm} for the particular choice \eqref{DH} of annihilation and creationlike variables. Adapting the derivation to the family of fermionic variables considered here, we get
\begin{equation}
\label{schrofermi}
i\partial_{\phi}\psi_D(\mathcal{N}_D,\phi)=\frac{l_{0}\langle \widehat{V^{2/3}e^{\tilde{\alpha}}\tilde{H}_D}  \rangle_{\Gamma}-C_D^{(\Gamma)}(\phi)}{ \langle \hat{\tilde{\mathcal H}}_0 \rangle_{\Gamma}}\,\psi_D(\mathcal{N}_D,\phi)\equiv {\mathcal H}_D^{(\Gamma)}(\phi)\psi_D(\mathcal{N}_D,\phi) .
\end{equation}
Here, the hat over classical observables indicates their corresponding representation as operators, according to the prescriptions of the works that we have already mentioned. Besides, the brackets $\langle.\rangle_{\Gamma}$ stand for the expectation value in $\Gamma$, taken with respect to the inner product in $\mathcal{H}_\text{kin}^\text{grav}$. Since the momentum of $\phi$ does not appear in $\tilde{H}_D$, the right-hand side of \eqref{schrofermi} represents a $\phi$-dependent operator (or a family of operators labeled by $\phi$, as one prefers) acting  on the fermionic sector. Hence, one may interpret this operator as the ({\emph {effective}}) Hamiltonian that generates the evolution of the fermionic degrees of freedom in the time $\phi$. This Hamiltonian, ${\mathcal H}_D^{(\Gamma)}(\phi)$, captures the most relevant features of the quantum background spacetime by means of the expectation values on $\Gamma$ and the specific quantum representation of the geometry that is employed. 

On the other hand, the function $C_D^{(\Gamma)}(\phi)$, added to similar contributions that arise from the scalar and tensor perturbations, provides the mean value in $\Gamma$ of the difference between $(\hat{\tilde{\mathcal H}}_0)^2+\partial_{\phi}^2$ and the Hamiltonian constraint of the unperturbed model\footnote{We notice here a typo in Ref. \cite{hybr-ferm}, where $C_D^{(\Gamma)}(\phi)$ and the rest of the backreaction contributions in Eqs. (6.5)-(6.7) of that paper should appear divided by $ \langle \hat{\tilde{\mathcal H}}_0 \rangle_{\Gamma}$.} \cite{hybr-ferm}. Thus, it can be understood as the fermionic contribution to the quantum backreaction on the homogeneous background, inasmuch as the mentioned difference actually measures how much $\Gamma$ departs from an exact solution of the unperturbed system.

Since the term $\tilde{H}_D$ is a sum over all possible fermionic modes, the Schr\"odinger equation \eqref{schrofermi} may be decomposed in a collection of individual equations, one for each of the modes. The fermionic contribution to the backreaction, $C_D^{(\Gamma)}(\phi)$ then depends on the behavior of the mode solutions. In fact, one does not always get a well-defined fermionic backreaction without applying regularization techniques. This issue critically depends on the asymptotic tail of the fermionic Hamiltonian ${\mathcal H}_D^{(\Gamma)}(\phi)$, when expressed as a sum over modes. And therefore it depends on the set of annihilation and creationlike variables chosen to describe the fermionic degrees of freedom. Thus, in order to analyze the possible divergence of $C_D^{(\Gamma)}(\phi)$, we study the solutions to Eq. \eqref{schrofermi}. In doing this, it is most convenient to view the Hamiltonian ${\mathcal H}_D^{(\Gamma)}(\phi)$ as the generator of some Heisenberg-like dynamics for the fermionic annihilation and creation operators. In fact, from Eq. \eqref{schrofermi} one can easily get the associated Heisenberg equations, taking into account the decomposition of $\tilde{H}_D$ as a sum over modes of the functions $\tilde{H}_{\vec{k}}$, that have an asymptotic behavior determined by Eq. \eqref{tildeHk}. In more detail, if we introduce the following state-dependent change to a conformal time
\begin{equation}
\label{etaGamm}
d{\eta_\Gamma}=  \frac{l_0\langle \hat{V}^{2/3} \rangle_{\Gamma} }{ \langle \hat{\tilde{\mathcal H}}_0 \rangle_{\Gamma} } d\phi,
\end{equation}
which is well-defined thanks to the the positivity of $\hat{\tilde{\mathcal H}}_0$ and the lower positive bound on the volume in each superselection sector of loop quantum cosmology, we obtain the following Heisenberg equations, evaluated at $\eta_{\Gamma}=\eta$:
\begin{eqnarray}
\label{Heisen}
d_{\eta_\Gamma}{\hat a}_{\vec{k}}^{(x,y)}(\eta,\eta_0)&=&- i F_{\vec{k}}^{(\Gamma)}{\hat a}_{\vec{k}}^{(x,y)}(\eta,\eta_0)+ G_{\vec{k}}^{(\Gamma)} {\hat b}_{\vec{k}}^{(x,y)\dagger}(\eta,\eta_0),\nonumber
\\
d_{\eta_\Gamma}{\hat b}_{\vec{k}}^{(x,y)\dagger}(\eta,\eta_0)&=& i \left(F_{\vec{k}}^{(\Gamma)}+\tilde{J}_{\vec{k}}^{(\Gamma)}\right){\hat b}_{\vec{k}}^{(x,y)\dagger}(\eta,\eta_0)- \bar{G}_{\vec{k}}^{(\Gamma)} {\hat a}_{\vec{k}}^{(x,y)}(\eta,\eta_0),
\end{eqnarray}
where, in the asymptotic regime of large $\omega_k$,
\begin{eqnarray}
\label{JkGamm}
\tilde{J}_{\vec{k}}^{(\Gamma)}&=&\frac{\langle  \widehat{2e^{\tilde{\alpha}}V^{2/3}h_{J}^{\vec{k}}} \rangle_{\Gamma} }{\langle \hat{V}^{2/3} \rangle_{\Gamma}},
\\
\label{FkGamm}
F_{\vec{k}}^{(\Gamma)}&=&\frac{\langle \widehat{ V^{2/3}\xi_k}\rangle_{\Gamma}+2\langle \widehat{e^{\tilde{\alpha}}V^{2/3}h_D^{\vec{k}}} \rangle_{\Gamma} }{\langle \hat{V}^{2/3} \rangle_{\Gamma}},
\\
\label{GkGamm}
G_{\vec{k}}^{(\Gamma)}&=& \frac{2i\omega_k\langle \widehat{ e^{i(F_{2}^{\vec{k},(x,y)}-J_{\vec{k}}^{(x,y)})}V^{2/3}  {\theta}_{\vec{k}}^{(x,y)}}\rangle_{\Gamma} +i\langle\widehat{ e^{i(F_{2}^{\vec{k},(x,y)}-J_{\vec{k}}^{(x,y)})}V^{2/3}{h}_{I}^{\vec{k}}}\rangle_{\Gamma}} {\langle \hat{V}^{2/3} \rangle_{\Gamma}}.
\end{eqnarray}
The factors $h_{J}^{\vec{k}}$, $h_{D}^{\vec{k}}$, and $h_{I}^{\vec{k}}$ are given in Eqs. \eqref{hj}-\eqref{hi2}. Provided that our prescriptions for the representation of the homogeneous geometry promote real functions to (at least) symmetric operators, we have that $F_{\vec{k}}^{(\Gamma)}$ and $\tilde{J}_{\vec{k}}^{(\Gamma)}$ are real. In addition, we assume that the state $\Gamma$ is such that all the considered functions admit asymptotic expansions in the limit of infinitely large $\omega_k$. The coefficients of these expansions are expectation values in $\Gamma$ of mode-independent operators. Actually, for our discussion, it suffices that the expansions exist up to terms of the order of a certain inverse power of $\omega_k$.

The Heisenberg equations determine a family of annihilation and creation operators parametrized by different values $\eta$ of $\eta_{\Gamma}$, once one fixes as initial data at $\eta_{\Gamma}=\eta_0$ the annihilation and creation operators ${\hat a}_{\vec{k}}^{(x,y)}$ and ${\hat b}_{\vec{k}}^{(x,y)\dagger}$ that appear in the fermionic Hamiltonian (together with their adjoints). It is straightforward to see that each such family of operators, ${\hat a}_{\vec{k}}^{(x,y)}(\eta,\eta_0)$ and ${\hat b}_{\vec{k}}^{(x,y)\dagger}(\eta,\eta_0)$, can be obtained by means of a Bogoliubov transformation from the initial ones, ${\hat a}_{\vec{k}}^{(x,y)}$ and ${\hat b}_{\vec{k}}^{(x,y)\dagger}$. In order to analyze the properties of that transformation, we follow a strategy that is close to the one developed in Ref. \cite{hybr-ferm} for the particular choice of variables \eqref{DH}. In the present and more general case, nonetheless, the analysis has some peculiarities that affect the asymptotic regime of large $\omega_k$. So, let us study in detail this asymptotic behavior. 

We first introduce the following fermionic operators, motivated in part by the previous definitions \eqref{ab}-\eqref{ftwo} of the annihilation and creationlike variables and by the dominant asymptotic term in $F_{\vec{k}}^{(\Gamma)}$,
\begin{eqnarray}
\label{qxy}
{\hat x}_{\vec{k}}(\eta,\eta_0)&=& f_{1,k}^{(\Gamma)}{\hat a}_{\vec{k}}^{(x,y)}(\eta,\eta_0)+ e^{-i \int^\eta_{\eta_0} d\eta_{\Gamma}\,\tilde{J}_{\vec{k}}^{(\Gamma)}}f_{2,k}^{(\Gamma)} {\hat b}_{\vec{k}}^{(x,y)\dagger}(\eta,\eta_0),
\\\nonumber
{\hat y}_{-\vec{k}-2\vec{\tau}}^{\dagger}(\eta,\eta_0)&=& f_{2,k}^{(\Gamma)}{\hat a}_{\vec{k}}^{(x,y)}(\eta,\eta_0)- e^{-i \int^\eta_{\eta_0} d\eta_{\Gamma}\,\tilde{J}_{\vec{k}}^{(\Gamma)}}f_{1,k}^{(\Gamma)} {\hat b}_{\vec{k}}^{(x,y)\dagger}(\eta,\eta_0),
\end{eqnarray}
where
\begin{equation}
\label{quantumDEHf}
f_{1,k}^{(\Gamma)}= \sqrt{ \frac{ \tilde{F}_k^{(\Gamma)} - \omega_k }{ 2 \tilde{F}_k^{(\Gamma)} } } , \qquad f_{2,k}^{(\Gamma)} = \sqrt{  \frac{ \tilde{F}_k^{(\Gamma)} + \omega_k }{ 2 \tilde{F}_k^{(\Gamma)} } },\qquad \tilde{F}_k^{(\Gamma)}=\frac{\langle \widehat{ V^{2/3}\xi_k}\rangle_{\Gamma}}{\langle \hat{V}^{2/3} \rangle_{\Gamma}}.
\end{equation}
Notice that $f_{1,k}^{(\Gamma)}$ and $f_{2,k}^{(\Gamma)}$ are both real functions for sufficiently large $\omega_k$, given the asymptotic behavior of $\xi_k$, and they satisfy $|f_{1,k}^{(\Gamma)}|^2+|f_{2,k}^{(\Gamma)}|^2=1$. These newly introduced operators inherit the following dynamics from Eq. \eqref{Heisen}:
\begin{eqnarray}
\label{eomx}
d_{\eta_\Gamma}{\hat x}_{\vec{k}}(\eta,\eta_0)&=& i \left[\omega_k\bigg(1+\frac{{F}_{\vec{k}}^{(\Gamma)}-\tilde{F}_k^{(\Gamma)}}{\tilde{F}_k^{(\Gamma)}}\bigg)+P^{(\Gamma)}_{\vec{k}}\right] {\hat x}_{\vec{k}}(\eta,\eta_0)+ H_{\vec{k}}^{(\Gamma)} {\hat y}_{-\vec{k}-2\vec{\tau}}^{\dagger}(\eta,\eta_0),
\\
\nonumber
d_{\eta_\Gamma}{\hat y}_{-\vec{k}-2\vec{\tau}}^{\dagger}(\eta,\eta_0)&=& -i \left[\omega_k\bigg(1+\frac{{F}_{\vec{k}}^{(\Gamma)}-\tilde{F}_k^{(\Gamma)}}{\tilde{F}_k^{(\Gamma)}}\bigg)+P^{(\Gamma)}_{\vec{k}}\right]
{\hat y}_{-\vec{k}-2\vec{\tau}}^{\dagger}(\eta,\eta_0) - \bar{H}_{\vec{k}}^{(\Gamma)} {\hat x}_{\vec{k}}(\eta,\eta_0),
\end{eqnarray}
with the definitions
\begin{eqnarray}\label{PGamma}
P^{(\Gamma)}_{\vec{k}}&=&\frac{\sqrt{\big(\tilde{F}_k^{(\Gamma)}\big)^2-\omega_k^2}}{\tilde{F}_k^{(\Gamma)}}\Im\left(G_{\vec{k}}^{(\Gamma)}e^{i\int^{\eta}_{\eta_0} d\eta_{\Gamma}\,\tilde{J}_{\vec{k}}^{(\Gamma)}}\right),
\\
\label{HGamma}
H_{\vec{k}}^{(\Gamma)} &=& - G_{\vec{k}}^{(\Gamma)}e^{i\int^{\eta}_{\eta_0} d\eta_{\Gamma}\,\tilde{J}_{\vec{k}}^{(\Gamma)}} - i \sqrt{\big(\tilde{F}_k^{(\Gamma)}\big)^2-\omega_k^2}\left(1+\frac{{F}_{\vec{k}}^{(\Gamma)}-\tilde{F}_k^{(\Gamma)}}{\tilde{F}_k^{(\Gamma)}}\right) +\frac{ \omega_k \Big( {\tilde{F}}_{\vec{k}}^{(\Gamma)} \Big)^{\prime}} { 2 \tilde{F}_k^{(\Gamma)} \sqrt{\big(\tilde{F}_k^{(\Gamma)}\big)^2-\omega_k^2}}+iQ_{\vec{k}}^{(\Gamma)},\\
\label{QGamma}
Q_{\vec{k}}^{(\Gamma)}&=&\frac{\tilde{F}_k^{(\Gamma)}+\omega_k}{\tilde{F}_k^{(\Gamma)}}\Im\left(G_{\vec{k}}^{(\Gamma)}e^{i\int^{\eta}_{\eta_0} d\eta_{\Gamma}\,\tilde{J}_{\vec{k}}^{(\Gamma)}}\right).
\end{eqnarray}
Here, the prime denotes the derivative with respect to $\eta_{\Gamma}$ and $\Im(.)$ is the imaginary part. Employing now the compact notation $\{\hat{z}_{\vec{k}}\}=\{{\hat x}_{\vec{k}},{\hat y}_{-\vec{k}-2\vec{\tau}}\}$ and introducing the rescaled operators $\hat{\tilde{z}}_{\vec{k}}=(iH_{\vec{k}}^{(\Gamma)})^{-1/2} \hat{z}_{\vec{k}}$, these all turn out to satisfy the same second order equation:
\begin{equation}\label{zeq}
\hat{\tilde{z}}_{\vec{k}}^{\prime\prime}=- \bigg[ \tilde{\omega}_{\vec{k}}^2 + \Big|H_{\vec{k}}^{(\Gamma)}\Big|^2 -i\tilde{\omega}_{\vec{k}}^{\prime}\bigg] \hat{\tilde{z}}_{\vec{k}},
\end{equation}
where
\begin{equation}\label{tildeo}
\tilde{\omega}_{\vec{k}}=\omega_k\bigg(1+\frac{{F}_{\vec{k}}^{(\Gamma)}-\tilde{F}_k^{(\Gamma)}}{\tilde{F}_k^{(\Gamma)}}\bigg)+P^{(\Gamma)}_{\vec{k}} + \frac{i}{2} \Big(\ln{H_{\vec{k}}^{(\Gamma)}}\Big)^{\prime}.
\end{equation}
It can be checked that two independent solutions of the linear differential equation \eqref{zeq} are $\tilde{z}^{l}_{\vec{k}}=\exp[-i(-1)^{l}\tilde\Theta^{l}_{\vec{k}}]$ with
\begin{equation}\label{tildez}
\tilde\Theta^{l}_{\vec{k}}(\eta_0)=0, \quad (\tilde\Theta^{l}_{\vec{k}})^{\prime}=\tilde{\omega}_{\vec{k}}+\Lambda^{l}_{\vec{k}},\qquad l=1,2,
\end{equation}
where $\Lambda^{l}_{\vec{k}}$ are the solutions of the Ricatti equation
\begin{eqnarray}
\label{lambdaeq}
\Big( \Lambda_{\vec{k}}^l \Big)^{\prime} &=& i (-1)^l  \Big[ \Big(\Lambda_{\vec{k}}^l \Big)^2 +2 \tilde{\omega}_{\vec{k}}\Lambda_{\vec{k}}^l \Big] - u^l_{\vec{k}}, \\
\label{ul}
u^l_{\vec{k}} &=& i (-1)^l  | H_{\vec{k}}^{(\Gamma)} |^2 + \Big[(-1)^l+1\Big]\tilde{\omega}_{\vec{k}}^{\prime},
\end{eqnarray}
with initial conditions $\Lambda^{l}_{\vec{k}}(\eta_0)=0$. The corresponding independent solutions for $\hat{z}_{\vec{k}}$, after undoing the scaling, are then given by $z^{l}_{\vec{k}}=\exp[-i(-1)^{l}\Theta^{l}_{\vec{k}}]$, where
\begin{equation}
\label{Theta}
\Theta_{\vec{k}}^l=\omega_k(\eta-\eta_0)+\frac{i}{2} \Big[(-1)^l+1\Big]  \ln{\left( \frac{H_{\vec{k}}^{(\Gamma)}} {H_{\vec{k}}^{(\Gamma),0} } \right) }  +\omega_k\int_{\eta_0}^{\eta} d\eta_{\Gamma}\, \left(\frac{{F}_{\vec{k}}^{(\Gamma)}-\tilde{F}_k^{(\Gamma)}}{\tilde{F}_k^{(\Gamma)}}\right)+\int_{\eta_0}^{\eta} d\eta_{\Gamma}\, \left(\Lambda_{\vec{k}}^{l}+ P_{\vec{k}}^{(\Gamma)}\right).
\end{equation}
From now on, we use a superindex or a subindex $0$ (on occasions preceded by a coma) to denote evaluation at $\eta_{\Gamma}=\eta_0$. With the above independent solutions of the second order equation at hand, the relation between ${\hat x}_{\vec{k}}$ and ${\hat y}_{-\vec{k}-2\vec{\tau}}^{\dagger}$ (or their adjoints) implied by the first order equations \eqref{eomx}, and the relation of these operators with the annihilation and creation operators in the Heisenberg picture, we can readily derive the dynamical Bogoliubov transformation of the latter as
\begin{eqnarray}
\label{qBogoliubov}
{\hat a}_{\vec{k}}^{(x,y)}(\eta,\eta_0)&=& \alpha_{\vec{k}}(\eta,\eta_0){\hat a}_{\vec{k}}^{(x,y)}+\beta_{\vec{k}}(\eta,\eta_0){\hat b}_{\vec{k}}^{(x,y)\dagger},\nonumber\\
{\hat b}_{\vec{k}}^{(x,y)\dagger}(\eta,\eta_0)&=&-e^{i\int_{\eta_0}^{\eta} d\eta_{\Gamma}\,\tilde{J}_{\vec{k}}^{(\Gamma)}}\bar{\beta}_{\vec{k}}(\eta,\eta_0){\hat a}_{\vec{k}}^{(x,y)}+e^{i\int_{\eta_0}^{\eta} d\eta_{\Gamma}\,\tilde{J}_{\vec{k}}^{(\Gamma)}}\bar{\alpha}_{\vec{k}}(\eta,\eta_0){\hat b}_{\vec{k}}^{(x,y)\dagger},
\end{eqnarray}
where the alpha and beta coefficients take the expressions
\begin{align}
\label{alphaq}
\alpha_{\vec{k}}&=
\bigg[f_{1,k}^{(\Gamma)} \Big(f_{1,k}^{(\Gamma),0}-f_{2,k}^{(\Gamma),0} \zeta_{\vec{k}}\Big) e^{i\int_{\eta_0}^{\eta} d\eta_{\Gamma}\, \tilde{\Lambda}_{\vec{k}}^1} - f_{2,k}^{(\Gamma)}f_{1,k}^{(\Gamma),0} \bar{\zeta}_{\vec{k}} \frac{\bar{H}_{\vec{k}}^{(\Gamma)}}{\bar{H}_{\vec{k}}^{(\Gamma),0}} e^{i\int_{\eta_0}^{\eta} d\eta_{\Gamma}\,\bar{\tilde{\Lambda}}_{\vec{k}}^2}\bigg] e^{i\omega_k\left[\eta-\eta_0+\int_{\eta_0}^{\eta} d\eta_{\Gamma}\, \frac{{F}_{\vec{k}}^{(\Gamma)}-\tilde{F}_k^{(\Gamma)}}{\tilde{F}_k^{(\Gamma)}}\right]}\nonumber\\  &+
\bigg[f_{2,k}^{(\Gamma)}\Big(f_{1,k}^{(\Gamma),0}\bar{\zeta}_{\vec{k}}+f_{2,k}^{(\Gamma),0} \Big) e^{-i\int_{\eta_0}^{\eta} d\eta_{\Gamma}\, \bar{\tilde{\Lambda}}_{\vec{k}}^1} + f_{1,k}^{(\Gamma)}f_{2,k}^{(\Gamma),0} \zeta_{\vec{k}} \frac{H_{\vec{k}}^{(\Gamma)}}{H_{\vec{k}}^{(\Gamma),0}} e^{-i\int_{\eta_0}^{\eta} d\eta_{\Gamma}\, \tilde{\Lambda}_{\vec{k}}^2}\bigg] e^{-i\omega_k\left[\eta-\eta_0+\int_{\eta_0}^{\eta} d\eta_{\Gamma}\, \frac{{F}_{\vec{k}}^{(\Gamma)}-\tilde{F}_k^{(\Gamma)}}{\tilde{F}_k^{(\Gamma)}}\right]},
\end{align}
\begin{align}\label{betaq}
\beta_{\vec{k}}&= \bigg[f_{1,k}^{(\Gamma)}\Big(f_{2,k}^{(\Gamma),0}+f_{1,k}^{(\Gamma),0} \zeta_{\vec{k}}\Big) e^{i\int_{\eta_0}^{\eta} d\eta_{\Gamma}\, \tilde{\Lambda}_{\vec{k}}^1} - f_{2,k}^{(\Gamma)} f_{2,k}^{(\Gamma),0} \bar{\zeta}_{\vec{k}} \frac{\bar{H}_{\vec{k}}^{(\Gamma)}}{\bar{H}_{\vec{k}}^{(\Gamma),0}} e^{i \int_{\eta_0}^{\eta} d\eta_{\Gamma}\,\bar{\tilde{\Lambda}}_{\vec{k}}^2}\bigg] e^{i\omega_k\left[\eta-\eta_0+\int_{\eta_0}^{\eta} d\eta_{\Gamma}\, \frac{{F}_{\vec{k}}^{(\Gamma)}-\tilde{F}_k^{(\Gamma)}}{\tilde{F}_k^{(\Gamma)}}\right]}\nonumber\\  &+
\bigg[f_{2,k}^{(\Gamma)}\Big(f_{2,k}^{(\Gamma),0}\bar{\zeta}_{\vec{k}}-f_{1,k}^{(\Gamma),0} \Big) e^{-i\int_{\eta_0}^{\eta} d\eta_{\Gamma}\, \bar{\tilde{\Lambda}}_{\vec{k}}^1} - f_{1,k}^{(\Gamma)}f_{1,k}^{(\Gamma),0} \zeta_{\vec{k}} \frac{H_{\vec{k}}^{(\Gamma)}}{H_{\vec{k}}^{(\Gamma),0}} e^{-i\int_{\eta_0}^{\eta} d\eta_{\Gamma}\, \tilde{\Lambda}_{\vec{k}}^2}\bigg] e^{-i\omega_k\left[\eta-\eta_0+\int_{\eta_0}^{\eta} d\eta_{\Gamma}\, \frac{{F}_{\vec{k}}^{(\Gamma)}-\tilde{F}_k^{(\Gamma)}}{\tilde{F}_k^{(\Gamma)}}\right]}.
\end{align}
Here, we have defined
\begin{align}
\label{zetaq}
\zeta_{\vec{k}}&=  \frac{iH_{\vec{k}}^{(\Gamma),0}}{2\omega_k\Big[1+\left({F}_{\vec{k}}^{(\Gamma),0}-\tilde{F}_k^{(\Gamma),0}\right)/\tilde{F}_k^{(\Gamma),0}\Big]+i\Big(\ln{H_{\vec{k}}^{(\Gamma)}}\Big)^{\prime}_0 + 2P_{\vec{k}}^{(\Gamma),0}},\\  \tilde{\Lambda}_{\vec{k}}^l&=\Lambda_{\vec{k}}^{l}+ P_{\vec{k}}^{(\Gamma)}.
\end{align}

\subsection{Unitarity and backreaction}

The Bogoliubov transformation of the annihilation and creation operators that implements the Heisenberg dynamics dictated by Eq. \eqref{Heisen} may be used to obtain solutions of the associated Schr\"odinger equation \eqref{schrofermi} \cite{hybr-ferm}. In order to do so, nonetheless, it is necessary that the transformation admits a unitary implementation in the fermionic Fock space $\mathcal{F}_D$, for all values of initial and final times, $\eta_0$ and $\eta$. If this is the case, one can construct the unitary operator that integrates the Heisenberg equation. Evolving with it the Fock vacuum defined by the initial operators $\{{\hat a}_{\vec{k}}^{(x,y)},{\hat b}_{\vec{k}}^{(x,y)}\}$, one indeed arrives at a solution of the Schr\"odinger equation. Other solutions can be similarly found starting with the initial $n$-particle states.

Actually, the considered Bogoliubov transformation is unitarily implementable in Fock space if and only if the sequence  $\{\beta_{\vec{k}}(\eta,\eta_0)\}_{\vec{k}\in\mathbb{Z}^{3}}$ is square summable \cite{shale,dere}. This summability exclusively depends on the asymptotic behavior of the beta coefficients, in the regime of large $\omega_k$, provided that they are regular in their dependence on $\eta$ and $\eta_0$ for all $\vec{k}\in\mathbb{Z}^3$. This should be the case with the adopted loop representation of the homogeneous geometry (and suitable operator prescriptions), assuming that $\theta_{\vec{k}}^{(x,y)}$ is taken as a smooth function of the geometric degrees of freedom. Therefore, we are interested in the asymptotic behavior of all functions and quantities appearing in Eq. \eqref{betaq}. On the one hand, as it was shown in Ref. \cite{hybr-ferm}, we have 
\begin{equation}
\label{FkGammaser}
\tilde{F}_k^{(\Gamma)}=\omega_k+    \frac{M^{2}}{2l_0^{2}\omega_k}  W_1^{(\Gamma)}+{\mathcal{O}}(\omega_k^{-3}),\qquad
W_1^{(\Gamma)}= \frac{\langle \hat{V}^{4/3}\rangle_{\Gamma}} {\langle \hat{V}^{2/3}\rangle_{\Gamma}},
\end{equation}
where we recall that $M$ is the bare mass of the Dirac field. Then,
\begin{equation}
\label{f12ser}
f_{1,k}^{(\Gamma)}= \frac{M}{2l_0\omega_k} \sqrt{ W_1^{(\Gamma)} }+ {\mathcal{O}}(\omega_k^{-3}),\qquad
f_{2,k}^{(\Gamma)}= 1- \frac{M^{2}}{8l_0^2\omega_k^2} W_1^{(\Gamma) }+ {\mathcal{O}}(\omega_k^{-4}).
\end{equation}
On the other hand, from the asymptotic behavior of $h_{D}^{\vec{k}}=\mathcal{O}(1)$ and $h_{I}^{\vec{k}}$, that follows from Eqs. \eqref{hd}-\eqref{hi2} together with condition \eqref{dcond}, we get
\begin{equation}
\frac{{F}_{\vec{k}}^{(\Gamma)}-\tilde{F}_k^{(\Gamma)}}{\tilde{F}_k^{(\Gamma)}}=\frac{2\langle \widehat{e^{\tilde{\alpha}}V^{2/3}h_D^{\vec{k}}} \rangle_{\Gamma} }{\langle \widehat{{V}^{2/3}\xi_k} \rangle_{\Gamma}}=\mathcal{O}(\omega_k^{-1}),\qquad P_{\vec{k}}^{(\Gamma)}=\mathcal{O}\left(G_{\vec{k}}^{(\Gamma)}\omega_k^{-1}\right)=\mathcal{O}\left(\mathrm{Max}[\omega_k^{-2},\theta^{(x,y)}_{\vec{k}}]\right),
\end{equation}
and therefore
\begin{equation}
\label{HkGammaser}
H_{\vec{k}}^{(\Gamma)}= - \bar{G}_{\vec{k}}^{(\Gamma)}e^{-i \int^{\eta}_{\eta_0} d\eta_{\Gamma}\,\tilde{J}_{\vec{k}}^{(\Gamma)}}+\frac{M}{ l_0} \sqrt{ W_1^{(\Gamma)}  }  \left[ -i\left(1+\frac{2\langle \widehat{e^{\tilde{\alpha}}V^{2/3}h_D^{\vec{k}}} \rangle_{\Gamma} }{\langle \widehat{{V}^{2/3}\xi_k} \rangle_{\Gamma}}\right) + \frac{1}{4\omega_k} \left(\ln{  W_1^{(\Gamma)}   } \right)^{\prime\,}\, \right]
+{\mathcal{O}}\left(\mathrm{Max}[\omega_k^{-2},{G}_{\vec{k}}^{(\Gamma)}\omega_k^{-2}]\right),
\end{equation}
so that
\begin{equation}
\label{zetaser}
\zeta_{\vec{k}}=\frac{M}{2l_0\omega_k} \sqrt{ W_1^{(\Gamma),0} }  
-\frac{i}{2\omega_k}\bar{G}_{\vec{k}}^{(\Gamma),0}+{\mathcal{O}}\left(\mathrm{Max}[\omega_k^{-m},{G}_{\vec{k}}^{(\Gamma)}\omega_k^{-2}]\right),
\end{equation}
where $m=2$ for $\vec{k}\in\mathbb{Z}^3_{\uparrow}$, whereas $m=3$ for $\vec{k}\in\tilde{\mathbb{Z}}^3$. The remaining functions that we have to analyze in order to derive the asymptotic behavior of $\beta_{\vec{k}}(\eta,\eta_0)$ are the solutions $\Lambda^{l}_{\vec{k}}$ of the Ricatti equation \eqref{lambdaeq}. Their behavior depends drastically on the function $u_{\vec{k}}^{l}$, given in Eq. \eqref{ul}. It is not difficult to see that, provided condition \eqref{dcond} holds for second order derivatives, all the contributions to those functions are of asymptotic order $\mathcal{O}(1)$, except possibly for $| H_{\vec{k}}^{(\Gamma)} |^2$. For this specific quantity, a look at Eq. \eqref{HkGammaser} reveals that one gets a contribution that may grow as $\omega_k^2(\theta^{(x,y)}_{\vec{k}})^2$. In particular, it is $\mathcal{O}(1)$ if $\theta^{(x,y)}_{\vec{k}}=\mathcal{O}(\omega_k^{-1})$. Recalling the characterization of the possible asymptotic behavior allowed for $\theta^{(x,y)}_{\vec{k}}$, described in the previous section, we have the following scenarios:
\begin{itemize}
	\item[a)] For $\vec{k}\in\tilde{\mathbb{Z}}^{3}$ or $\vec{k}\in{\mathbb{Z}}^{3}_{\uparrow,1}\subset{\mathbb{Z}}^{3}_{\uparrow}$, with  $\theta^{(x,y)}_{\vec{k}}=\mathcal{O}(\omega_k^{-1})$ in ${\mathbb{Z}}^{3}_{\uparrow,1}$, the source term $u_{\vec{k}}^{l}$ of the Ricatti equation \eqref{lambdaeq} is asymptotically $\mathcal{O}(1)$ and the solutions $\Lambda^{l}_{\vec{k}}$ satisfy
	\begin{equation}\label{Lambdaasympt1}
	\int_{\eta_0}^{\eta} d\eta_{\Gamma}\, \Lambda_{\vec{k}}^{l} =
	- (-1)^l\frac{i }{2\omega_k}  \int_{\eta_0}^{\eta} d\eta_{\Gamma}\,u^{l}_{\vec{k}}   + \mathcal{O}(\omega_{\vec{k}}^{-2})=\mathcal{O}(\omega_k^{-1}),
	\end{equation}
	similarly as it happened in Ref. \cite{hybr-ferm}. This can be checked by solving Eq. \eqref{lambdaeq}, with vanishing initial condition  and after ignoring the nonlinear term, by means of a repeated integration by parts [taking into account condition \eqref{dcond}]. With the result, one can estimate the order of the ignored term, obtaining Eq. \eqref{Lambdaasympt1}. 
	\item[b)] For tuples $\vec{k}$ in the complement (up to a finite subset) ${\mathbb{Z}}^{3}_{\uparrow,2}$ of ${\mathbb{Z}}^{3}_{\uparrow,1}$ in ${\mathbb{Z}}^{3}_{\uparrow}$, that is such that $\omega_k^{-1}=o\left(\theta^{(x,y)}_{\vec{k}}\right)$, we use that $u^{l}_{\vec{k}}=\mathcal{O}\left(\omega_k^2[\theta^{(x,y)}_{\vec{k}}]^2\right)$. Let us notice, however, that it is only the imaginary part of $u^{l}_{\vec{k}}$ that gives a growing contribution in the asymptotic regime of large $\omega_k$, as can be seen from definition \eqref{ul}.  It follows that one may again compute the solutions of the linear part of the Ricatti equation \eqref{lambdaeq}, that we call $\lambda_{\vec{k}}^{l}$, with vanishing initial conditions. In this way, one finds 
	\begin{equation}\label{lambdaasympt}
	\int_{\eta_0}^{\eta} d\eta_{\Gamma}\, \Re(\lambda_{\vec{k}}^{l}) =
	\frac{1 }{2\omega_k}  \int_{\eta_0}^{\eta} d\eta_{\Gamma}\,| H_{\vec{k}}^{(\Gamma)} |^2  + o(1),\qquad \int_{\eta_0}^{\eta} d\eta_{\Gamma}\, \Im(\lambda_{\vec{k}}^{l}) = o(1),
	\end{equation}
	where $\Re(.)$ is the real part. Taking into account this behavior, and iteratively repeating the same analysis for the subdominant contributions to $\lambda_{\vec{k}}^{l}$ in the solution $\Lambda_{\vec{k}}^{l}$ of the entire Ricatti equation, one can show that, asymptotically,
	\begin{equation}\label{Lambdaasympt2}
	\int_{\eta_0}^{\eta} d\eta_{\Gamma}\, \Re(\Lambda_{\vec{k}}^{l}) =
	\gamma_{\vec{k}}  + o(1),\qquad \int_{\eta_0}^{\eta} d\eta_{\Gamma}\, \Im(\Lambda_{\vec{k}}^{l}) = o(1),
	\end{equation} 
	where $\gamma_{\vec{k}}$ does not depend on $l$. Therefore, in particular, in the asymptotic regime of large $\omega_k$,
	\begin{equation}
	\quad e^{- i(-1)^{l}\int_{\eta_0}^{\eta} d\eta_{\Gamma}\,{\Lambda}_{\vec{k}}^l}=e^{- i(-1)^{l}\int_{\eta_0}^{\eta} d\eta_{\Gamma}\,\Re({\Lambda}_{\vec{k}}^l)}[1+o(1)].
	\end{equation}
\end{itemize}

Employing all this asymptotic information, we can easily show that the alpha and beta coefficients \eqref{alphaq} and \eqref{betaq}  have the following behavior for infinitely large $\omega_k$:
\begin{align}\label{alphaser1}
&\alpha_{\vec{k}}(\eta,\eta_0)= e^{-i\omega_k(\eta-\eta_0)-i\int_{\eta_0}^{\eta} d\eta_{\Gamma}\, \Xi_{1}^{\vec{k}}}+o(1),\\\label{betaser1}
&\beta_{\vec{k}}(\eta,\eta_0)= \frac{i}{2\omega_k}\bigg\{G_{\vec{k}}^{(\Gamma),0}e^{-i\omega_k(\eta-\eta_0)-i\int_{\eta_0}^{\eta} d\eta_{\Gamma}\, \Xi_{1}^{\vec{k}}}-
G_{\vec{k}}^{(\Gamma)}e^{i\omega_k(\eta-\eta_0)+i\int_{\eta_0}^{\eta}d\eta_{\Gamma}\,[\tilde{J}_{\vec{k}}^{(\Gamma)}+\Xi_{2}^{\vec{k}}]}\bigg\}+\delta_{\vec{k}},\\ \label{deltakas}
&\delta_{\vec{k}}={\mathcal{O}}\left(\mathrm{Max}[\omega_k^{-3},{G}_{\vec{k}}^{(\Gamma)}\omega_k^{-2}]\right) \quad {\rm when} \quad \vec{k}\in\tilde{\mathbb{Z}}^3,\qquad \delta_{\vec{k}}=o(G_{\vec{k}}^{(\Gamma)}\omega_k^{-1}) \quad {\rm when} \quad \vec{k}\in\mathbb{Z}^3_{\uparrow},
\end{align}
where we have defined, with $l=1,2$,
\begin{equation}
\Xi_{l}^{\vec{k}}=\omega_k\frac{{F}_{\vec{k}}^{(\Gamma)}-\tilde{F}_k^{(\Gamma)}}{\tilde{F}_k^{(\Gamma)}}\quad \mathrm{for}\quad \vec{k}\in\tilde{\mathbb{Z}}^3\cup\mathbb{Z}^3_{\uparrow,1},\qquad\Xi_{l}^{\vec{k}}=\omega_k\frac{{F}_{\vec{k}}^{(\Gamma)}-\tilde{F}_k^{(\Gamma)}}{\tilde{F}_k^{(\Gamma)}}+\Re(\Lambda^{l}_{\vec{k}})\quad \mathrm{for}\quad \vec{k}\in\mathbb{Z}^3_{\uparrow,2}.
\end{equation}
Then, in all cases, we have that
\begin{equation}
\beta_{\vec{k}}(\eta,\eta_0)=\mathcal{O}\left(\mathrm{Max}[\omega_k^{-3},G_{\vec{k}}^{(\Gamma)}\omega_k^{-1}]\right).
\end{equation} 
Since the sequences that define any of the two quantities in the $\mathrm{Max}$ function are square summable over $\mathbb{Z}^{3}$ [see the definition of $G_{\vec{k}}^{(\Gamma)}$, together with the asymptotic expression for $h_I^{\vec{k}}$ and condition \eqref{unitf1}], we can conclude that the transformations implied by the Heisenberg equations \eqref{Heisen} are unitarily implementable in Fock space.

A comment is in order at this point. In our previous analysis, we have assumed that $H_{\vec{k}}^{(\Gamma)}\neq 0$. If this were not the case, it is not hard to convince oneself that the beta coefficients of the dynamical Bogoliubov transformation would be of the same asymptotic order as $\omega_k^{-1}$, given the behavior of $f_{1,k}^{(\Gamma)}$ and $f_{2,k}^{(\Gamma)}$. However, from Eq. \eqref{HkGammaser} one can check that, for $H_{\vec{k}}^{(\Gamma)}$ to vanish, $\theta^{(x,y)}_{\vec{k}}$ must be precisely of order $\mathcal{O}(\omega_k^{-1})$. Since $\theta^{(x,y)}_{\vec{k}}$ forms a square summable sequence by assumption, that might only happen for $\vec{k}$ in some subset of $\mathbb{Z}^3_{\uparrow}$ where any sequence that is $\mathcal{O}(\omega_k^{-1})$ turned out to be square summable. Therefore, the Heisenberg dynamics would also be unitarily implementable in this particular case. Taking this into account, our following analysis about the backreaction can be applied to all possible scenarios.

Once we have confirmed the unitarity of the Heisenberg dynamics determined by our quantum expectation values over the homogeneous geometry, which do not even need to correspond to a background described by effective loop quantum cosmology, we can proceed to construct solutions of the associated Schr\"odinger equation \eqref{schrofermi} by evolving the initial Fock vacuum with the corresponding unitary operator. In order to do so, we follow the strategy of Ref. \cite{hybr-ferm}, conveniently generalized to the present situation but avoiding the repetition of redundant computations. First of all, given the asymptotic formula \eqref{alphaser1}, it is most convenient to split the operator that implements the Heisenberg dynamics into the composition of two unitaries. The first one incorporates the dominant $\eta_{\Gamma}$-dependent phase of the alpha coefficients, namely, it is the unitary operator associated with the Bogoliubov transformation
\begin{equation}
{\hat a}_{\vec{k}}^{(x,y)}\longrightarrow e^{-i\omega_k(\eta-\eta_0)-i\int_{\eta_0}^{\eta} d\eta_{\Gamma}\, \Xi_{1}^{\vec{k}}}{\hat a}_{\vec{k}}^{(x,y)},\qquad
{\hat b}_{\vec{k}}^{(x,y)\dagger}\longrightarrow e^{i\omega_k(\eta-\eta_0)+i\int_{\eta_0}^{\eta} d\eta_{\Gamma}\,[\tilde{J}_{\vec{k}}^{(\Gamma)}+\Xi_{1}^{\vec{k}}]}{\hat b}_{\vec{k}}^{(x,y)\dagger}.
\end{equation}
The second unitary operator then completes the dynamical transformation \eqref{qBogoliubov} by implementing the linear mapping
\begin{eqnarray}
\label{tildeBog}
{\hat a}_{\vec{k}}^{(x,y)}&\longrightarrow& \tilde{\alpha}_{\vec{k}}(\eta,\eta_0){\hat a}_{\vec{k}}^{(x,y)}+\tilde{\beta}_{\vec{k}}(\eta,\eta_0){\hat b}_{\vec{k}}^{(x,y)\dagger},\nonumber\\
{\hat b}_{\vec{k}}^{(x,y)\dagger}&\longrightarrow&-\bar{\tilde{\beta}}_{\vec{k}}(\eta,\eta_0){\hat a}_{\vec{k}}^{(x,y)}+\bar{\tilde{\alpha}}_{\vec{k}}(\eta,\eta_0){\hat b}_{\vec{k}}^{(x,y)\dagger},
\end{eqnarray}
with
\begin{equation}
\label{tildealphabeta}
{\tilde \alpha}_k(\eta,\eta_0)=e^{i\omega_k(\eta-\eta_0)+i\int_{\eta_0}^{\eta} d\eta_{\Gamma}\, \Xi_{1}^{\vec{k}}}\alpha_k(\eta,\eta_0),\qquad   {\tilde \beta}_k(\eta,\eta_0)=e^{-i\omega_k(\eta-\eta_0)-i\int_{\eta_0}^{\eta}d\eta_{\Gamma}\,[\tilde{J}_{\vec{k}}^{(\Gamma)}+\Xi_{1}^{\vec{k}}]}\beta_k(\eta,\eta_0).
\end{equation}
This latter operator can be written in the form $e^{-\hat{T}}$, with \cite{hybr-ferm}
\begin{eqnarray}
\label{quantumT}
\hat{T}&=&\sum_{\vec{k}\neq\vec{\tau},(x,y)} \!\!\Big[ \Delta_{\vec{k}} {\hat a}_{\vec{k}}^{(x,y)\dagger} {\hat b}_{\vec{k}}^{(x,y)\dagger}  - \bar{\Delta}_{\vec{k}} {\hat b}_{\vec{k}}^{(x,y)} {\hat a}_{\vec{k}}^{(x,y)}
- i \rho_{\vec{k}} \Big( {\hat a}_{\vec{k}}^{(x,y)\dagger} {\hat a}_{\vec{k}}^{(x,y)} +  {\hat b}_{\vec{k}}^{(x,y)\dagger}  {\hat b}_{\vec{k}}^{(x,y)} \Big) + i c_{\vec{k}}^{(x,y)} \Big],
\end{eqnarray}
where $c_{\vec{k}}^{(x,y)}\in\mathbb{R}$ is an undetermined phase, and we have chosen the following parametrization of the (modified) Bogoliubov coefficients:
\begin{eqnarray}
\label{alphaparametrized}
{\tilde \alpha}_{\vec{k}}=\cos{\sqrt{|\Delta_{\vec{k}}|^2+\rho_{\vec{k}}^2}} +i \rho_{\vec{k}} \frac{\sin{\sqrt{|\Delta_{\vec{k}}|^2+\rho_{\vec{k}}^2}}}{\sqrt{|\Delta_{\vec{k}}|^2+\rho_{\vec{k}}^2}},\qquad 
{\tilde \beta}_{\vec{k}}=-\Delta_{\vec{k}} \frac{\sin{\sqrt{|\Delta_{\vec{k}}|^2+\rho_{\vec{k}}^2}}}{\sqrt{|\Delta_{\vec{k}}|^2+\rho_{\vec{k}}^2}}.
\end{eqnarray}

An analogous calculation to that presented in Ref. \cite{hybr-ferm}, taking now due care of the additional phase contribution $\tilde{J}_{\vec{k}}^{(\Gamma)}$, shows that the evolution of the fermionic Fock vacuum by the combined action of the two introduced unitary operators indeed gives rise to solutions of the Schr\"odinger equation \eqref{schrofermi}, provided that
\begin{equation}
\label{backreaction}
C_D^{(\Gamma)}(\phi)=l_0\langle \hat{V}^{2/3}\rangle_{\Gamma} \sum_{\vec{k},(x,y)} \Big[ \Im{(G_{\vec{k}}^{(\Gamma)}\bar{\Delta}_{\vec{k}})} - d_{\eta_{\Gamma}}c_{\vec{k}}^{(x,y)}\Big].
\end{equation} 
This is the fermionic contribution to the backreaction. In the rest of this section, we analyze the convergence of this fermionic backreaction by using our asymptotic analyses above. In fact, one might always set the quantity $C_D^{(\Gamma)}$ equal to 0 by means of an appropriate choice of $c_{\vec{k}}^{(x,y)}$, i.e., by conveniently tuning the phase of the solutions $\psi_D$ to the Schr\"odinger equation, even if the total sum of these phases could then diverge. Ignoring this fine-tuning of the phases, and hence avoiding the possible resummation of two individually divergent quantities, we focus our attention on the terms that depend on $\bar{\Delta}_{\vec{k}}$. From our previous definitions and considerations, it is not difficult to check that ${\tilde \alpha}_{\vec{k}}=1+o(1)$ in the asymptotic regime of large $\omega_k$. Therefore, the parametrization \eqref{alphaparametrized} implies that the asymptotically dominant term in $\Delta_{\vec{k}}$ is the same as for ${\tilde \beta}_{\vec{k}}$. Using Eq. \eqref{betaser1} and the asymptotic behavior of $\Re(\Lambda^{l}_{\vec{k}})$ shown in Eq. \eqref{Lambdaasympt2}, we then obtain
\begin{align}\label{brasymp}
\Im{(G_{\vec{k}}^{(\Gamma)}\bar{\Delta}_{\vec{k}})} =& \frac{1}{2\omega_k}\bigg\{|G_{\vec{k}}^{(\Gamma)}|^{2}-\Re[{G}_{\vec{k}}^{(\Gamma)}\bar{G}_{\vec{k}}^{(\Gamma),0}]\cos\left[2\omega_k(\eta-\eta_0)+\int_{\eta_0}^{\eta} d\eta_{\Gamma}\,(\tilde{J}_{\vec{k}}^{(\Gamma)}+2\Xi_{1}^{\vec{k}})\right]\\\nonumber&+\Im[{G}_{\vec{k}}^{(\Gamma)}\bar{G}_{\vec{k}}^{(\Gamma),0}]\sin\left[2\omega_k(\eta-\eta_0)+\int_{\eta_0}^{\eta} d\eta_{\Gamma}\,(\tilde{J}_{\vec{k}}^{(\Gamma)}+2\Xi_{1}^{\vec{k}})\right]\bigg\}+\tilde{\delta}_{\vec{k}},
\end{align}
where the subdominant terms $\tilde{\delta}_{\vec{k}}$ are of the asymptotic order of $\delta_{\vec{k}}G_{\vec{k}}^{(\Gamma)}$, with the behavior of $\delta_{\vec{k}}$ being given in Eq. \eqref{deltakas}. Hence, to ensure that the backreaction is finite, without the need of introducing a divergent phase in the fermionic part of the states, we only have to impose that the sum over $\vec{k}\in\mathbb{Z}^3$ of the contributions in Eq. \eqref{brasymp} be absolutely convergent. In particular, this condition eliminates any ambiguity that might affect the nonabsolute sum, given the possibility of attaining conditional convergences. Besides, we naturally require that this contribution to the backreaction is well defined independently of the choice of homogeneous state $\Gamma$, and for all times $\eta$. Taking into account the different asymptotic behaviors allowed for $\theta_{\vec{k}}^{(x,y)}$, we contemplate the following cases:
\begin{itemize}
	\item[i)] For tuples $\vec{k}\in\mathbb{Z}^{3}_{\uparrow}$, we have from Eqs. \eqref{tildeHk} and \eqref{hi1} that $G_{\vec{k}}^{(\Gamma)}$ is of the same order as $\omega_k\theta_{\vec{k}}^{(x,y)}$. The subdominant term $\tilde{\delta}_{\vec{k}}$ in Eq. \eqref{brasymp} is then asymptotically negligible compared to $[G_{\vec{k}}^{(\Gamma)}]^2\omega_k^{-1}$, since $\theta_{\vec{k}}^{(x,y)}$ is of order $\omega_k^{-3/2}$ or higher in this case. Besides, with our assumptions [including condition \eqref{dcond}], the time-dependent oscillations in Eq. \eqref{brasymp} cannot be compensated, at dominant order, with the first term. Hence, we conclude that the contribution to the backreaction is absolutely summable over the considered modes, independently of $\Gamma$, if and only if the sequence $\{\omega_k|\theta_{\vec{k}}^{(x,y)}|^{2}\}_{\vec{k}\in\mathbb{Z}^{3}_{\uparrow}}$ is summable (regardless of the values of the canonical variables for the homogeneous geometry on which $\theta_{\vec{k}}^{(x,y)}$ may depend). The sufficiency of this condition for the oscillating terms in Eq. \eqref{brasymp} follows, in particular, from the use of the Cauchy-Schwarz inequality.
	
	\item[ii)] On the other hand, for $\vec{k}\in\tilde{\mathbb{Z}}^{3}$, we recall that $\omega_k\theta_{\vec{k}}^{(x,y)}$ must be negligible compared to $\omega_k^{-1/2}$. Employing the asymptotic expressions \eqref{tildeHk} and \eqref{hi2}, we conclude that $G_{\vec{k}}^{(\Gamma)}$ is either of the same order as $\omega_k\theta_{\vec{k}}^{(x,y)}$ or of order $\omega_k^{-1}$, whichever is dominant, unless these two types of contributions are of the same order and cancel each other. If this cancellation did not happen, at least for a nonempty infinite subset $\tilde{\mathbb{Z}}^{3}_{1}\subseteq\tilde{\mathbb{Z}}^{3}$, it is not difficult to realize that the terms \eqref{brasymp} would not be absolutely summable over $\tilde{\mathbb{Z}}^{3}$, given the asymptotic growth of the density of states with the Dirac eigenvalue $\omega_k$. Therefore, it is necessary that the term $2\omega_k\theta_{\vec{k}}^{(x,y)}$ in $\tilde{H}_{\vec{k}}$ cancels any possible contribution of order $\omega_k^{-1}$ in $h_I^{\vec{k}}$, up to terms that are $o(\omega_k^{-1})$. Imposing this requirement, and recalling condition \eqref{dcond}, we must have that, for $\vec{k}\in\tilde{\mathbb{Z}}^{3}_{1}$, 
	\begin{eqnarray}\label{brcond1}
	\theta_{\vec{k}}^{(x,y)}=-i\frac{\tilde{M}e^{-{\alpha}}}{4\omega_k^{2}}{\pi}_{\alpha}e^{iF^{\vec{k},(x,y)}_2}+\vartheta_{\vec{k}}^{(x,y)},
	\end{eqnarray}
	where $\vartheta_{\vec{k}}^{(x,y)}=o(\omega_k^{-2})$. This is a necessary condition for the absolute convergence of the terms \eqref{brasymp} in  $\tilde{\mathbb{Z}}^{3}$. Inserting this behavior into the interacting part of $\tilde{H}_{\vec{k}}$, and considering its relation with $G_{\vec{k}}^{(\Gamma)}$, one can show that this latter quantity has the same asymptotic order, for $\vec{k}\in\tilde{\mathbb{Z}}^{3}_{1}$, as the dominant contribution among the terms $\omega_k\vartheta_{\vec{k}}^{(x,y)}$ and $\omega_k^{-2}$. The latter type of term automatically provides, when introduced in Eq. \eqref{brasymp}, a convergent series in $\tilde{\mathbb{Z}}^{3}_{1}$. Thus, following analogous arguments to those explained in our previous case, we reach the conclusion that the sufficient condition in $\tilde{\mathbb{Z}}^{3}$ for the absolute convergence of the considered fermionic backreaction is that
	\begin{equation}\label{brcond2}
	\sum_{\vec{k}\in\tilde{\mathbb{Z}}^{3}_{1}}\omega_k|\vartheta_{\vec{k}}^{(x,y)}|^{2}<\infty
	\end{equation}
	and that the sequence $\{\mathrm{Max}[\omega_k^{-3},\omega_k|\theta_{\vec{k}}^{(x,y)}|^{2}]\}_{\vec{k}\in\tilde{\mathbb{Z}}^{3}_{2}}$ be summable if the complement $\tilde{\mathbb{Z}}^{3}_{2}$ of $\tilde{\mathbb{Z}}^{3}_{1}$ in $\tilde{\mathbb{Z}}^{3}$ is infinite.
\end{itemize}

All of these conditions, that ensure that the backreaction contribution $C_D^{(\Gamma)}$ is well defined without introducing any regularization scheme, impose much more severe ultraviolet restrictions to the choice of fermionic annihilation and creationlike variables than the unitarity requirement \eqref{unitf1}. Besides, it is worth emphasizing that the asymptotic behavior characterized by conditions \eqref{brcond1} and \eqref{brcond2} must hold for $\vec{k}$ in a nonempty infinite subset $\tilde{\mathbb{Z}}^{3}_{1}$ of the lattice ${\mathbb{Z}}^{3}$, while each of the subsets for which one must demand the rest of conditions stated in the cases i and ii above might be empty. At the end of the day, the asymptotic behavior of the characteristic density of states of the Dirac eigenvalues in $T^3$ determines the specific form of these conditions. Because of this, if we further restricted the choice of annihilation and creationlike variables (e.g. by symmetry considerations) so that they could not depend on the tuple $\vec{k}$ except through the corresponding eigenvalue $\omega_k$, we would conclude that the studied fermionic backreaction would be absolutely convergent if and only if conditions \eqref{brcond1} and \eqref{brcond2} are asymptotically satisfied for all $\vec{k}\in\mathbb{Z}^{3}$ (except, possibly, a finite subset).

Finally, let us comment that one may want to restrict even further the choice of fermionic variables in order to guarantee that the Hamiltonian operator that appears in the Schr\"odinger equation \eqref{schrofermi} has a well-defined action on the Fock vacuum. As a consequence, the Hamiltonian would then be properly defined in the dense subset of the Fock space $\mathcal{F}_D$ spanned by the $n$-particle/antiparticle states, that have a finite number of fermionic excitations. In that case, the constraint equation (and thus the Schr\"odinger equations derived from it) would indeed be a rigorously defined equation, at least in what concerns the fermionic degrees of freedom. Given the normal ordering adopted in the fermionic Hamiltonian, it is clear that only the interacting terms, that annihilate and create infinite pairs of particles and antiparticles, may prevent the image of the vacuum providing a normalizable state in $\mathcal{F}_D$. In fact, this nomalizability holds if and only if the terms that multiply $a_{\vec{k}}^{(x,y)}b_{\vec{k}}^{(x,y)}$ (and their complex conjugates) in the decomposition of $\tilde{H}_D$ as a sum over modes form a square summable sequence. Arguments like those that we have explained show that this happens if and only if one imposes conditions that are similar to the ones displayed in i-ii above, but demanding the stronger requirement of the summability of the sequences
\begin{equation} \{\omega^2_k|\theta_{\vec{k}}^{(x,y)}|^{2}\}_{\vec{k}\in\mathbb{Z}^{3}_{\uparrow}},\quad \{\omega^2_k|\vartheta_{\vec{k}}^{(x,y)}|^{2}\}_{\vec{k}\in\tilde{\mathbb{Z}}^{3}_{1}}, \quad {\rm and} \quad \{\mathrm{Max}[\omega_k^{-2},\omega^2_k|\theta_{\vec{k}}^{(x,y)}|^{2}]\}_{\vec{k}\in\tilde{\mathbb{Z}}^{3}_{2}}.
\end{equation}

\section{Conclusions}

In this work, we have investigated a possible procedure to avoid some of the typical divergences of quantum field theory in the context of hybrid loop quantum cosmology. Specifically, we have studied in detail the case of a Dirac field minimally coupled to an inflationary cosmology. The Dirac field has been treated as a perturbation, including its zero-mode if one exists, and in general additional scalar and tensor perturbations have been permitted. In our perturbative scheme, the action of the system is truncated at second order in all the perturbations. After decomposing the inhomogeneities in suitable modes, defined on the spatial hypersurfaces of the homogeneous model, the resulting relativistic system is subject to the zero mode of the Hamiltonian constraint, that in particular contains all the relevant fermionic contribution to the Hamiltonian, as well as to (an infinite mode collection of) perturbative constraints that are linear in the metric and scalar perturbations. At the considered quadratic perturbative order of our truncation, the time-dependent mode coefficients that describe the fermionic field are automatically gauge invariant with respect to these perturbative constraints. Besides, the rest of the perturbations in the system can be described by means of a set of canonical variables that are formed by the well-known Mukhanov-Sasaki and tensor gauge invariants, and by an Abelianized version of the linear perturbative constraints, together with all their momenta. The hybrid approach for the quantization of this cosmological system is based in a convenient Fock representation for each of the perturbative sectors of the phase space, combined with a less standard quantum gravity-inspired representation of the purely homogeneous degrees of freedom (that can be thought to describe an inflationary FLRW cosmology on their own), namely the representation employed in loop quantum cosmology.

We have focused our analysis on divergences that may arise in the quantum theory from the standard Fock treatment of the fermionic degrees of freedom. Actually, we have explored the possibility of avoiding that these infinities appear by taking into consideration the fact that it is the whole phase space of the cosmological system what has to be treated quantum mechanically in a hybrid way, rather than only the fermionic degrees of freedom, while the FLRW cosmology is maintained as a classical entity. As commented above, this means that each sector of the total phase space is given a qualitatively different quantum representation. This applies in particular to what one may call the homogeneous background sector and the Dirac perturbations. Within this context, it does not seem unnatural to question whether one may separate them in different ways, and thus assign different dynamical roles to each of these sectors. These different alternatives for the splitting can be realized in practice, without affecting the rest of scalar and tensor perturbations, by considering canonical transformations of the fermionic variables that depend on the homogeneous background. When these transformations are completed to be canonical for the entire system (at the considered perturbative order of our truncation), the Hamiltonian that generates the dynamics of the new fermionic variables changes with respect to the original one. We are then tempted to expect that, with an adequate splitting of the joint dynamics of the geometric FLRW degrees of freedom and the Dirac field, we may attain a satisfactory control of the divergences that arise from the quantum field theory representation of the fermionic variables in their corresponding Hamiltonian.

In more detail, here we have incorporated the freedom that exists in identifying the Heisenberg dynamics of the fermionic degrees of freedom, exploiting the different dynamical roles of the homogeneous background and of the fermionic perturbations, by introducing families of annihilation and creationlike variables that are obtained through background-dependent canonical transformations. The specific form of these transformations is {\emph{a priori}} only restricted by the following physical consideration [and a mild condition on their dependence on the homogeneous degrees of freedom: see Eq. \eqref{dcond}]. They must define variables that, in the context of quantum field theory in classical curved spacetimes, possess a nontrivial dynamics that is unitarily implementable in Fock space. Besides, the associated Fock vacuum must be invariant under the classical symmetries of the Dirac-FLRW system, and define a standard convention for particles and antiparticles. These families of annihilation and creationlike variables turn out to determine unitarily equivalent Fock representations of the Dirac field. With such a generic collection of different descriptions for the fermionic degrees of freedom, we have computed the form of the resulting Hamiltonian that generates their dynamics. In particular, we have characterized its asymptotic tail, when it is expressed as an infinite sum in terms of the annihilation and creation coefficients of the spatial eigenmodes of the Dirac operator. This fermionic Hamiltonian has a nontrivial dependence on the resulting homogeneous sector of the cosmological model. In fact, after implementing the hybrid quantization procedure and adopting a kind of Born-Oppenheimer ansatz for the physical quantum states that are annihilated by the zero mode of the entire constraint, one arrives at a fermionic Hamiltonian that is defined by means of expectation values over the homogeneous geometry, and that can be understood to generate a Schr\"odinger dynamics for the part of the states that encodes the information about the fermionic degrees of freedom. This Hamiltonian operator, which varies with the specific choice of annihilation and creationlike variables, generalizes the operator that would be obtained in quantum field theory on curved spacetimes, inasmuch as its dependence on the homogeneous background is not longer evaluated on a classical geometry, but replaced with the corresponding expectation values. We have carried out an asymptotic analysis, in the regime of large eigenvalues $\omega_k$ of the spatial Dirac operator, of the Heisenberg dynamics associated with this fermionic Hamiltonian, and we have shown that it amounts to a Bogoliubov transformation of the annihilation and creation operators which is unitarily implementable in Fock space. The vacuum state, when evolved with the corresponding unitary operator, can then be seen to provide solutions of the Schr\"odinger equation with a very specific backreaction term that depends on the geometric expectation values that define the considered dynamics. This backreaction term can serve to measure (in mean value) how much the homogeneous part of the quantum states departs from an exact solution of the unperturbed inflationary model. With the obtained asymptotic information about the Heisenberg dynamics, we have been able to characterize the choices of annihilation and creationlike variables, in the family under consideration, that allow for a finite fermionic backreaction, without the need of introducing any regularization technique or resummations of infinities based on a conditional convergence. Once we have guaranteed that the analyzed backreaction is well defined, we have seen that, with some slightly more stringent conditions on our choice of fermionic variables, we can also ensure that the fermionic Hamiltonian is actually a rigorously defined operator in the dense subset of the Fock space spanned by $n$-particle/antiparticle states.

The relevance of our characterization of the families of fermionic annihilation and creationlike variables that prevent divergences within the hybrid framework of loop quantum cosmology can be seen twofold. On the one hand, we have shown that, for an infinite number of modes, the canonical transformation that defines the fermionic variables must display a very specific asymptotic dependence on the homogeneous geometry, as well as on the mass of the Dirac field [see Eqs. \eqref{unitf1} and \eqref{brcond1}]. This dependence in the ultraviolet regime of large $\omega_k$ actually implies a severe restriction, in the quantum theory, about which part of the dynamical degrees of freedom of the system must be treated as geometric, and which part contains the information about the genuine fermionic excitations. On the other hand, it is clear that a specific characterization of the physically admissible annihilation and creationlike variables leads to a restriction on the choice of fermionic vacuum, among the infinitely many that are available (even when restricting all considerations to choices selected by the unitarity of the classical dynamics). In fact, the already mentioned, specific dependence on the homogeneous background of the transformations that define the fermionic variables has the effect of reducing the asymptotic order of the interaction terms in the corresponding fermionic Hamiltonian. One could think that a further restriction of the choice of fermionic variables, and therefore of their vacuum, is possible if one investigates even deeper the asymptotic tail of the Hamiltonian and tries to eliminate completely its interacting contribution. If this procedure were viable, the variables determined in this way for the description of the fermionic degrees of freedom would then diagonalize the resulting Hamiltonian, at least in the ultraviolet sector, and therefore might be thought to be optimally adapted to the quantum dynamics of the entire cosmological system.

Furthermore, the specification of suitable variables for the quantum description of the fermionic degrees of freedom can, at least in certain regimes of physical interest, shed light on the influence that the dynamics of this type of matter might have on the quantum evolution of the homogeneous background geometry. In this work, such effects can be found, first, in the redefinition of the scale factor and its momentum that is required in order that they remain canonical with respect to the introduced fermionic variables. Besides, at least at the level of expectation values, a genuinely quantum backreaction of the fermionic degrees of freedom on the behavior of the partial wave function that describes this homogeneous geometry is contained in the function $C_D^{(\Gamma)}$, inasmuch as it measures, in mean value, how much the homogeneous background differs from a quantum solution of the nonperturbed cosmology. These effects can be given precise formulas [see Eqs. \eqref{newscale}, \eqref{newpi} and \eqref{backreaction}] which, if shown to be well-defined quantities as $C_D^{(\Gamma)}$ has been seen to be here, can serve as a starting point for the quantitative determination of modifications that the presence of fermionic matter may introduce in the dynamics of the background geometry, with respect to the purely homogeneous scenario found in standard linearized cosmology (even when this is described within the context of loop quantum cosmology). These modifications would likely, in turn, leave some imprint in the evolution of the primordial perturbations of scalar and tensor type. Following techniques like those explored recently in Ref. \cite{hybrgui}, one may investigate the consequences and physical relevance that these modifications may have on the power spectrum of the cosmological perturbations, as well as on possible non-Gaussianities. Actually, it should be possible to perform \vline{an analysis similar to the one conducted} in this work in order to specify a privileged family of variables for the description of scalar and tensor perturbations in quantum cosmology, such that their quantum Hamiltonian and backreaction effects display well-behaved properties. If that were the case, it might even be possible to investigate the interplay between their associated backreaction functions [analogous to $C_D^{(\Gamma)}$] \cite{hybr-ferm}, the fermionic contribution, and the quantum evolution of the background geometry.

\acknowledgments

The authors are grateful to H. Sahlmann and T. Thiemann for enlightening conversations. This work was supported by Grants No. FIS2014-54800-C2-2-P and No. FIS2017-86497-C2-2-P from the Spanish MINECO (Ministerio de Economía y Competitividad).

\end{document}